\documentstyle[aps,preprint,eqsecnum,epsfig,rotate]{revtex}
\newcommand{\lapprox}{\raisebox{-0.5ex}{$\ 
\stackrel{\textstyle<}{\textstyle\sim}\ $}}
\tightenlines
\begin{document}
\draft
\title{Chiral Symmetry Restoration 
and Realisation of the Goldstone Mechanism
in the U(1) Gross-Neveu Model\\ 
at Non-Zero Chemical Potential}
\author{Ian Barbour$^a$,
Simon Hands$^b$, John B. Kogut$^c$, Maria-Paola Lombardo$^d$, 
Susan Morrison$^b$}
\address{$^a$ Department of Physics and Astronomy, \\
University of Glasgow,Glasgow G12 8QQ, U.K.\\}
\address{$^b$ Department of Physics, University of Wales Swansea,\\
Singleton Park, Swansea, SA2 8PP, U.K.}
\address{$^c$ Department of Physics,
University of Illinois at Urbana-Champaign\\
1110 West Green Street,
Urbana, IL 61801-3080, U.S.A.}
\address{$^d$ Istituto Nazionale di Fisica Nucleare, 
Laboratori Nazionali del Gran Sasso, \\I-67010 Assergi (AQ), Italy
\footnote{Current address} \\
and\\
Fakult\"at f\"ur Physik, Universit\"at Bielefeld, Postfach 100 31, 
D-33501 Bielefeld, Germany}
\maketitle
\begin{abstract}
We simulate the Gross-Neveu model in 2+1 dimensions at nonzero
baryon density (chemical potential $\mu\neq 0$).  
It is possible to formulate this model with a real action and therefore 
to perform 
standard hybrid Monte Carlo simulations
with $\mu\neq 0$ in the functional measure. We compare the physical observables from these simulations 
with simulations using
the Glasgow method where the value of $\mu$ in the functional
measure is fixed at a value $\mu_{upd}$. We find that the observables are
sensitive to the choice of $\mu_{upd}$. We consider the implications of
our findings for Glasgow method QCD simulations at $\mu\neq0$.
We demonstrate that the realisation of the Goldstone 
mechanism in the Gross-Neveu model is 
fundamentally different from that in QCD. We find that this difference
explains why there is an unphysical transition in QCD simulations at
$\mu\neq 0$ associated with the pion mass scale whereas the transition
in the Gross-Neveu model occurs at a larger mass scale and is therefore
consistent with theoretical predictions. We note classes of theories which
are exceptions to the Vafa-Witten theorem which permit the possibility
of formation of baryon number violating diquark condensates.  
\end{abstract}
\section{Introduction}

  Successful lattice simulations of QCD at nonzero baryon density
have yet to be achieved and the fundamental obstacle to success is
the fact that standard hybrid Monte Carlo techniques do not admit
the complex functional measure appropriate to the full theory. The results
of the quenched theory are known to be unphysical \cite{OLD,DAVKLEP,MISHA}. 
Attempts
have been made to study unquenched QCD at $\mu\neq0$ using the Glasgow 
method which involves generating a statistical ensemble via hybrid Monte Carlo
at $\mu=0$ , reweighting the observables by the ratio of the fermion determinant at $\mu\neq 0$ to
that at $\mu=0$ and performing a Grand Canonical Partition Function expansion in the fugacity variable 
$e^{\mu/T}$ to obtain observables for all $\mu$, but the results were not significantly different from 
the quenched theory. In this
paper we study the Gross-Neveu model which is one of the few models 
amenable to hybrid Monte Carlo simulations at nonzero density. By implementing
the Glasgow method in the Gross-Neveu model and comparing observables with
those of a standard hybrid Monte Carlo simulation we reveal the limitations
of the Glasgow method. The results are relevant to QCD simulations.
Work by Stephanov \cite{MISHA} strongly suggests that any nonzero density
simulation which incorporates a real path integral measure proportional
to $\det(MM^{\dagger})$ is doomed to failure due to the formation of
a light ``baryonic pion'' from a quark and a conjugate quark
\cite{DAVKLEP,Go88,KLS}. 
Since an 
earlier simulation \cite{HKK3} of the Gross-Neveu model at $\mu\neq 0$ did not
exhibit such a pathology an explanation is required. We provide a solution to
this puzzle later in this paper.

 Let us briefly review the current status of QCD simulations at nonzero 
density. In QCD the fermion determinant is complex for chemical potential $\mu$
 non-zero, therefore generating an ensemble by  Monte Carlo 
at $\mu \neq 0$  is impracticable. The fermion 
number density, $J_0=\langle\bar{\psi}\gamma_{0}\psi\rangle$, measures the excess of quarks relative to anti-quarks
and is expected to start to rise from zero at some value, $\mu_o$. This 
{\it onset}, $\mu_o$ corresponds to the point where the
phase of nuclear matter is more energetically favourable than the vacuum
state (for which $J_0\equiv 0$). Chiral symmetry restoration
occurs at some critical $\mu_c$ and we expect \cite{MISHA2} that $\mu_0\lapprox\mu_c\simeq m_p/3$, where $m_p$ is the proton mass. Quenched simulations, on the other hand,
predict 
$\mu_o \simeq m_{\pi}/2$ suggesting that in the
limit where the bare quark mass $m\rightarrow 0$ chiral symmetry is
restored for any $\mu\neq0$. This result is unphysical because the
pion which has baryon number zero, 
should not couple to the baryon chemical potential.

First attempts to simulate full QCD using
the Glasgow method \cite{BARBELL} also
produced perplexing results \cite{LAT96}.  
On an $8^4$ lattice with bare quark mass $m_b=0.01$ at $\beta=5.1$ we found  
$\mu_o \simeq 0.1$ which differs considerably from the strong coupling 
analysis \cite{BILIC}
prediction $\mu_c\simeq 0.65$ for $\beta=5.0$. Furthermore 
the scaling of $\mu_o$ with $m$ was consistent with a Goldstone
boson controlling the onset. This result has
motivated an assessment of the effectiveness of the Glasgow method
via its implementation in a simpler model.

Some insight into the effectiveness and the possible
problems  of the Glasgow method 
was obtained by a study of QCD  at infinite coupling \cite{SC}, where
we compared results obtained by use of the
Glasgow method  with the quenched   and the nearly exact ones
available in that limit. 
Glasgow results for $\mu \simeq m_\pi/2$  were found to reproduce
the quenched ones in quantitative detail. This rules out plausible 
physical explanations of the early Glasgow onset, which include 
anomalously low baryonic states or thermal excitations of baryons
\cite{AT92}\cite{BILIC}.  
In the present study the Glasgow results will be compared with an analogous 
of the quenched results (Toy model) and with the exact results of
\cite{HKK3}.

In section II we describe in detail the lattice formulation of the
3d GN model including a discussion of the 
symmetries and the features of the model which make it amenable to 
lattice simulation by the hybrid Monte Carlo algorithm even at nonzero
chemical potential. In section III we outline the Glasgow Method for
the GN model which makes use of eigenvalue symmetry relations to construct
a Grand Canonical Partition Function expansion in the fugacity variable.
The special feature of the Glasgow method is that in principle it permits 
simulations using any fixed value of $\mu$ in the functional measure
in order to obtain observables such as fermion number density for an 
extensive range of $\mu$ values ranging from $\mu=0$ to $\mu>\mu_c$. By
comparison with the hybrid Monte Carlo simulations where the 
$\mu\equiv\mu_{upd}$ in
the functional measure matches the 
$\mu\equiv\mu_{meas}$ at which the observable is 
calculated we assess the effectiveness of the Glasgow method within
the limits of our statistics. In section III we present the results of
our Gross-Neveu simulation using the Glasgow Method and provide a direct
comparison with results of a standard hybrid Monte Carlo simulation.
In section IV we present results of a Toy Model simulation of the GN
model. In the Toy Model the $\mu$ in the functional measure is different
from the $\mu$ at which the observable is calculated.
When $\mu$ in the functional measure is set to 0, the Toy model
is the Gross--Neveu equivalent of a quenched approximation.
When $\mu$ in the functional measure equals $\mu$ at which the 
observables are calculated, the Toy model is exact.
The Toy model results therefore give an insight into the relevance of
including the chemical potential into the dynamics. 
As discussed in \cite{SC}, the Glasgow method  should
improve over the quenched/Toy model because of the reweighting.
In the low statistics limit, the Glasgow method should reproduce
the Toy results, while in the Glasgow method
should approach the exact results. 
Hence the interest of considering both Toy and exact results
to understand the effectiveness of the Glasgow method.
In Section V of this paper we discuss the current onset, comparing
Toy, Glasgow, and HMC results. We will show that these onsets
are all controlled by the lowest excitation in the pseudoscalar
compostite spectrum. In contrast with QCD, this is not the Goldstone
pion, but a massive state. 
In section VI of this paper we consider a 
question which is not specific to the lattice simulation method but is 
highly relevant to our understanding of the physics of chiral symmetry 
restoration.
We provide evidence via a hybrid Monte Carlo simulation that the realisation
of the Goldstone mechanism in the Gross-Neveu model (specific to theories 
with four-fermion interactions) is fundamentally different from the familiar
mechanism in QCD. We show that the reason for this is that in
the Gross-Neveu model the dominant contributions to the Goldstone pion
come from the disconnected diagrams. This fact precludes
the existence of a light baryonic pion in $\mu\neq 0$ simulations 
incorporating four-fermion interactions. Finally in section VII we discuss the 
relevance of our Gross-Neveu model study to QCD simulations at nonzero
chemical potential.

\section{Lattice Formulation of the 3d GN model}
\label{sec:lat}

The fermionic part of the lattice action we have used 
for the bosonized Gross-Neveu model with U(1) chiral 
symmetry 
is given by \cite{HKK3}
\begin{eqnarray}
S_{fer} =  \bar\chi_i(x)M_{ijxy}\chi_j(y) 
  & = &\sum_{i=1}^{N}
\biggl[\sum_{x,y}\bar\chi_i(x){\cal M}_{x,y}\chi_i(y) \nonumber \\
  & + & 
{1\over8}\sum_x\bar\chi_i(x)\chi_i(x)
\Bigl(\sum_{<\tilde x,x>}\sigma(\tilde x) + i\varepsilon(x)
\sum_{<\tilde x,x>}
\pi(\tilde x)\Bigr)\biggr] 
\label{eq:Slat}
\end{eqnarray}
Here, $\chi_i$ and $\bar\chi_i$ are complex Grassmann-valued staggered
fermion fields defined on the lattice sites, the auxiliary scalar and
pseudoscalar fields $\sigma$ and $\pi$ are defined on the dual lattice
sites, and the symbol $\langle\tilde x,x\rangle$ denotes the set of 8 dual sites
$\tilde x$ adjacent to the direct lattice site $x$. $N$ is the number of
staggered fermion species and $1/g^{2}$ is the four-fermi coupling.
The symbol $\varepsilon(x)$ denotes the
alternating phase $(-1)^{x_0+x_1+x_2}$.
The auxiliary boson fields $\sigma$ and $\pi$ are weighted in the path integral 
by an additonal factor corresponding to
\begin{equation}
S_{aux}={N\over{2g^2}}\sum_{\tilde x}\sigma^2({\tilde x})+\pi^2({\tilde x}).
\end{equation}
The fermion kinetic operator ${\cal M}$ at non-zero density is given by
\begin{equation}
{\cal M}_{x,y} =  {1\over2}\Bigl[\delta_{y,x+\hat0}e^\mu-
\delta_{y,x-\hat0}e^{-\mu}\Bigr] 
+  {1\over2}\sum_{\nu=1,2}\eta_\nu(x)
\Bigl[\delta_{y,x+\hat\nu}-\delta_{y,x-\hat\nu}\Bigr] 
+ m\delta_{y,x},
\end{equation}
where $m$ is the bare fermion mass, $\mu$ is the chemical potential, and
$\eta_\nu(x)$ are the Kawamoto-Smit phases
$(-1)^{x_0+\cdots+x_{\nu-1}}$.

The model can be simulated using the hybrid Monte Carlo algorithm
\cite{DKPR}, in which complex bosonic pseudofermion fields $\Phi$ are updated
using the action $\Phi^\dagger(M^\dagger M)^{-1}\Phi$. After integration
over $\Phi$ the measure in the
functional integral is therefore manifestly real and given by
\begin{equation}
\mbox{det}(M^\dagger[\sigma,\pi] M[\sigma,\pi])\mbox{e}^{-S_{aux}[\sigma,\pi]}
=\mbox{det}({\cal M}+\sigma+i\varepsilon\pi)
\mbox{det}({\cal M}+\sigma-i\varepsilon\pi)\mbox{e}^{-S_{aux}[\sigma,\pi]},
\end{equation}
where we have used the fact the only complex entries of $M$ occur on the
diagonal. Therefore the simulation includes $N$ ``white'' flavors of fermion with 
positive axial charge (ie. coupling to $+i\pi$), and $N$ ``black'' flavors with
negative axial charge. If we include the global chiral symmetry valid for $m\rightarrow 0$
\begin{equation}
\chi(x)\mapsto\exp(i\alpha\varepsilon(x))\chi(x)\;\;\;;\;\;\;
\bar\chi(x)\mapsto\exp(i\alpha\varepsilon(x))\bar\chi(x)\;\;\;;\;\;\;
\phi\equiv(\sigma+i\pi)\mapsto\exp(-2i\alpha)\phi,
\label{eq:u1a}
\end{equation}
then we see that the model at non-zero lattice spacing 
has the symmetry:
${\rm U}(N)_V\otimes{\rm U}(N)_V\otimes{\rm U}(1)_A$.
It is the ${\rm U}(1)_A$ symmetry (\ref{eq:u1a}) which is
broken, either spontaneously by the dynamics of the system, or
explicitly by a bare fermion mass. In the continuum limit, it is possible to 
recast the model in terms of $N_f=4N$ flavors of four-component Dirac 
spinors $\psi,\bar\psi$ \cite{BB}, ie. $N_f/2$ white flavors and $N_f/2$ black;
the global symmetry group should enlarge to 
$\mbox{U}(N_f/2)_V\otimes\mbox{U}(N_f/2)_V\otimes\mbox{U}(1)_A$.
The continuum action approximated by (\ref{eq:Slat}) 
after integration over the auxiliary fields $\sigma$ and $\pi$ is
\cite{HK}:
\begin{eqnarray}
S_{cont} & = &  \sum_{p=1}^{N_f/2}\biggl[
\bar\psi^{white}_p({\partial\!\!\!/}+m)\psi^{white}_p+
\bar\psi^{black}_p({\partial\!\!\!/}+m)\psi^{black}_p \nonumber \\
& - & {2g^2\over{N_f}}\left[(\bar\psi^{white}_p\psi^{white}_p+
                            \bar\psi^{black}_p\psi^{black}_p)^2+
(\bar\psi^{white}_pi\gamma_5\psi^{white}_p-\bar\psi^{black}i\gamma_5
\psi^{black}_p)^2\right]\biggr].
\end{eqnarray}
                    
The relevant features of the 3d Gross-Neveu (GN) model for $\mu\neq 0$ studies 
are that it has a chiral transition with a massless pion in the broken phase
and it can be formulated such that the fermion determinant is 
positive definite for $\mu \neq 0$ \cite{HKK3}.

\section{Glasgow method for the 3d GN model}
 
The key feature of the Glasgow Method is that the chemical potential
$\mu_{upd}$ appearing in the functional measure is fixed to a constant
value while the associated observables are calculated at the appropriate
$\mu_{meas}$ which can be varied at will. The observables are calculated from
logarithmic derivatives of a  Grand Canonical Partition Function (GCPF) expansion in the fugacity variable $ e^{\mu / T}$ where $T$ is the temperature.
To compensate for generating the ensemble 
at fixed $\mu_{upd}$, the observables are ``reweighted'' by a ratio $R_{rw}$
of fermion determinants. This procedure of reweighting also serves to make 
the method exact i.e. the expression for the 
GCPF from which the observables are obtained is formally correct in the limit
of infinite statistics. In practice it is necessary to assess the effectiveness
of the Glasgow Method via direct comparison with a standard hybrid Monte Carlo
simulation in which $\mu_{upd}$ is variable and identical to the $\mu_{meas}$ appearing in the expression for the observable. 

Consider the  expression for the GCPF in the Gross-Neveu model: 
\begin{equation}
Z(\mu)=\int[d\sigma][d\pi]\,\det\left(M(\sigma,\pi,\mu,m)\right)\,e^{-S_{aux}}
\label{eqn:GCPF}
\end{equation}
where $M$ is the fermion determinant, $S_{aux}$ contains the auxiliary fields
$\sigma$ and $\pi$, $m$ is
the bare/current quark mass and $\mu$ is the chemical potential.
The GCPF (for fixed $m$) can be rescaled and expressed as an ensemble average of $\det M$ at some fixed value $\mu=\mu_{upd}$: 
\begin{eqnarray}
Z(\mu) = 
{{ 
\int [d\sigma][d\pi] \, 
{{\det M(\mu_{meas})}\over{\det M(\mu_{upd})}} 
\det M(\mu_{upd})\,e^{-S_{aux}}
}
\over 
{
\int {[d\sigma][d\pi]\,\det {M(\mu_{upd})\,e^{-S_{aux}}}}
}} 
= \left<
{{\det M(\mu_{meas})} \over {\det M(\mu_{upd})}} 
\right >{\biggr\vert}_{\mu_{upd}} 
\label{eqn:reweight}
\end{eqnarray}
where angled brackets denote an average over an ensemble generated with 
chemical potential $\mu_{upd}$, and $\mu_{meas}$ is the chemical potential for which the 
physical observables are to be calculated.
 Note that for example generating the ensemble at $\mu_{upd}=0$ would allow us to circumvent the problem of a  complex action in a Monte Carlo simulation. 

For optimum efficiency of 
the Glasgow method we require 
a large overlap between the ensemble generated using $\det M (\mu_{upd})$ in
the functional measure and the
exact ensemble generated using $\det M (\mu_{meas})$. Let us define a reweighting 
factor $R_{rw}\equiv{{\det M(\mu_{meas},m)} \over {\det M(\mu_{upd},m)}}$.
The relative magnitude of the factor $R_{rw}$ configuration by configuration
gives a measure of the overlap. If there is poor overlap between the simulated ensemble 
and the true ensemble it is conceivable that only a small fraction of the 
configurations will contribute significantly to $Z$ (those where
$R_{rw}$ is large in magnitude) in which case extremely
high statistics would be required to extract sensible results.

The full lattice action for the bosonized GN model with $U(1)$ chiral symmetry 
is given in (\ref{eq:Slat}). The functional measure used in the HMC algorithm is $\det(MM^{\dagger})$. To formulate the Glasgow Method for this model we
express the Dirac fermion matrix, $M$ and a related matrix ${\hat{M}}$,  
in terms of matrices $G$ and $V$ where $G$ contains
all the spacelike links while $V$ ($V^{\dagger}$) contains the forward(backward) timelike links. Note that for the Gross-Neveu model 
$\det(M{\hat{M}})=\det(MM^{\dagger})$.
\begin{equation}
       2iM_{xy}(\mu)=Y_{xy} + G_{xy} + V_{xy} e^{\mu} +V^{\dagger}_{xy} e^{-\mu} 
 \nonumber \;\;;\;\;
-2i{\hat{M}}_{xy}(\mu)=Y_{xy}^{\dagger} + G_{xy} + V_{xy} e^{\mu} +V_{xy}^{\dagger} e^{-\mu} \nonumber
\end{equation}
The term describing the Yukawa couplings of scalars 
to fermions is given (in terms of the auxiliary fields $\sigma$
and $\pi$ on dual lattice sites $\tilde x$) by
\begin{equation}
Y_{xy} = 2i( m + \frac{1}{8} \sum_{<x,\tilde x>} \left(\sigma( \tilde x)
                   +i\varepsilon\pi( \tilde x)\right))\delta_{xy}. 
\end{equation}

The determinants of these fermion matrices are related to that of
the propagator matrix $P$ (following Gibbs \cite{GIBBS2}):
\begin{equation}
P=\left(\begin{array}{cc}
-GV-YV & V \\
  -V   & 0
\end{array} \right)
\label{eqn:propmat}
\end{equation}

which is a matrix of dimension $2n_s^2 n_t$ in this model, for a $n_s^2 \times 
n_t$ lattice. Note that $V$ is an overall factor of $P$. The inverse of the 
propagator matrix is
\begin{equation}
P^{-1}=V^{\dagger} \left(\begin{array}{cc}
0 & -1\\
1 & -G-Y
\end{array} \right) \;\;\;{\rm and}\;\;\;
{(P^{-1})}^{\dagger}= \left(\begin{array}{cc}
0 & 1\\
-1 & -G-Y^{\dagger}
\end{array} \right)V
\end{equation}
Note that $Y^{\dagger}=Y^{*}$

The determinants of $P$ and $M$ are simply related:
\begin{eqnarray}
\det (P - e^{-\mu})&=& \left|\begin{array}{cc}
-GV-YV -e^{-\mu}  & V \\
  -V                & -1e^{-\mu}\end{array} \right| \nonumber \\
& = & \det \left(GVe^{-\mu} + YVe^{-\mu} +e^{-2\mu} +V^2\right) \nonumber \\
& = & \det \left(\left(Ge^{-\mu} + Ye^{-\mu} + V^{\dagger}e^{-2\mu} +V\right)V\right)
\nonumber \\
& = & e^{-\mu n_s^2n_t} \,\det\left(G + Y +
V^{\dagger}e^{-\mu}+Ve^{\mu}\right) \nonumber \\
& = & e^{-\mu n_s^2n_t} \,\det (2M)
\end{eqnarray}
where we have used $\det V=1$ and $V^\dagger V = V V^\dagger =1$.
Similarly the determinants of $\hat{P}$ and $\hat{M}$ are simply
related. 

 We shall expand $\det (M{\hat{M}})$ as a polynomial in $e^{n\mu}$
and in so doing we make use of two symmetries of the eigenvalues.
First consider the transformation 
$\Lambda=\left(\begin{array}{cc}
 0  & 1 \\
-1  & 0
\end{array} \right)$ in the space of the propagator matrix. One
can easily show that

\begin{eqnarray}
\Lambda{(P^{-1})}^{\dagger}\Lambda^{\dagger}=\left(\begin{array}{cc}
-GV-Y^{\dagger} & 1 \\
  -1   & 0
\end{array} \right)={\hat{P}}
\end{eqnarray}
Thus if $\lambda$ is an eigenvalue of $P$ then $\frac{1}{{\lambda}^*}$
is an eigenvalue of $\hat{P}$. The presence of ${\lambda}^*$ and 
$\frac{1}{\lambda}$ in the spectrum then follows from the fact that
$\det(M\hat{M})$ is real for arbitrary fugacity. 

We now perform our GCPF expansion:

\begin{eqnarray}
\det(M\hat{M})&=&e^{2{n_s}^{2}n_t\mu}\det(P-e^{-\mu})\det(\hat{P}-e^{-\mu}) \\
          &=&e^{2{n_s}^{2}n_t\mu}\prod_{i=1}^{{n_s}^{2}n_t}
\left(\lambda_{i}-e^{-\mu}\right)\left(\frac{1}{\lambda_{i}}-e^{-\mu}\right)
\left(\lambda_{i}^{*}-e^{-\mu}\right)\left(\frac{1}{\lambda_{i}^{*}}-e^{-\mu}\right)\\
         &=&e^{2{n_s}^{2}n_t\mu}\prod_{i=1}^{{n_s}^{2}n_t}
\left({1-e^{-\mu}(\lambda_{i}+\frac{1}{\lambda_{i}})+e^{-2\mu}}\right)
\left({1-e^{-\mu}(\lambda^{*}_{i}+\frac{1}{\lambda^{*}_{i}})+e^{-2\mu}}
\right)\\
&=&\prod_{i=1}^{{n_s}^{2}n_t}\left({e^{\mu}+e^{-\mu}+\lambda_{i}+
\frac{1}{\lambda_{i}}}\right)\left({e^{\mu}+e^{-\mu}+\lambda_{i}^{*}+
\frac{1}{\lambda_{i}^{*}}}\right)\\
&=&\sum_{n=0}^{2n_{s}^{2}n_t}a_{n}{(e^{\mu}+e^{-\mu})}^{n}
\end{eqnarray}

The above GCPF expansion incorporates all of the eigenvalue symmetries of the model.  Provided the $\left< a_n \right>$ are determined to sufficient accuracy,
 we can measure the averaged characteristic polynomial over the ensemble
generated at any fixed $\mu=\mu_{upd}$ using a hybrid Monte Carlo algorithm, and use this to provide an analytic 
continuation \cite{BARBELL} for the GCPF to any non-zero $\mu$.
In order to obtain the fugacity expansion we must determine the eigenvalues of $P\hat{P}$. 

Since the matrix $V$ is an overall factor of $P$, 
consider the effect of multiplying a timelike link by $e^{2\pi i}$. 
We can perform a unitary transform to spread this over all timelike links so 
that a new symmetry emerges: 
\begin{equation} 
V \longrightarrow V \times \:\mbox{element of} \: Z(n_t)
\end{equation}
This is then transferred to the eigenvalues, $\lambda$
\begin{equation}
\det (P- \lambda_{i})=0
\end{equation}
Therefore the eigenvalues themselves have a
$Z(n_t)$ symmetry. This $Z(n_t)$ symmetry holds configuration by configuration. As a consequence of this symmetry, the characteristic polynomial
for $P$ is a polynomial in $e^{\mu n_t}$ with $(4n_s^2+1)$ real coefficients. 
 Thus we obtain an expansion for the GCPF 
in the fugacity, $e^{\mu/T}$
\begin{equation}
Z\propto\sum_{n=0}^{2n_s^2}\langle a_{n}\rangle{\left({e^{\mu n_t}+e^{-\mu n_t}}\right)}^n\equiv\sum_{n=-2n_s^2}^{2n_s^2}e^{-(\epsilon_{n}-n\mu)/T}
\end{equation}
 The major computational task in performing the GCPF fugacity expansion
is the
determination of all of the eigenvalues of $P\hat{P}$. It is more efficient to diagonalize
$(P\hat{P})^{n_t}$ than to diagonalize $P\hat{P}$, therefore we exploit the $Z(n_t)$
symmetry of the eigenvalues 
. This introduces a $Z(n_t)$ degeneracy
of the eigenvalues and effectively reduces the dimension of the matrix to
$4n_s^2n_t$ to $4n_s^2$. Note that although $P\hat{P}$ is a sparse matrix  
$(P\hat{P})^{n_t}$
will be dense.
The expansion coefficients $\langle a_{n}\rangle$ are evaluated in the simulation
and thermodynamic observables can be obtained from derivatives of $\ln Z$. In 
particular the number density is defined by:

\begin{equation}
\langle J_{0}\left(\mu,ma\right)\rangle \equiv \lim_{n_{s}\rightarrow \infty} \left[\frac {T}{{n_s}^2}
\frac{\partial \ln(Z(\mu,m))}{\partial \mu}\right]
\end{equation}

and in terms of the GCPF expansion 

\begin{equation}
\left<J_{0}(\mu)\right>={\frac{\sum_{n=-2{n_s}^2}^{2{n_s}^2}n \,e^{-(\epsilon_n - n\mu)/T}}
{\sum_{n=-2{n_s}^2}^{2{n_s}^2}e^{-(\epsilon_n - n\mu)/T}}}.
\label{eqn:numden}
\end{equation}

Since the 3d GN $U(1)$ model has a positive definite functional measure 
we can choose any desired $\mu_{upd}$ and investigate the influence 
of this choice on the observables. 
Standard hybrid Monte Carlo (HMC) simulations \cite{HKK3} showed a clear separation of the Goldstone pion mass scale and the critical chemical potential 
($\mu_c$) where chiral symmetry was restored
i.e.  $\mu_c>>m_\pi/2$. Thus the lattice
Gross-Neveu model is an example of a theory where there is no manifestation of
 a light Goldstone pion carrying non-zero baryon number even when the model is simulated with a real path integral measure proportional to 
$\det (MM^{\dagger})$.

We set out to investigate whether by applying the  Glasgow Method for a given $\mu_{upd}$ we could obtain from the GCPF expansion coefficients a fermion number density as a function of $\mu$ which would match the results of the
standard hybrid Monte Carlo simulations. Of particular interest was whether
we would observe the discontinuity in the number density at 
$\mu_c\gg m_{\pi}/2$ via the Glasgow Method.

\begin{figure}
\epsfig{file=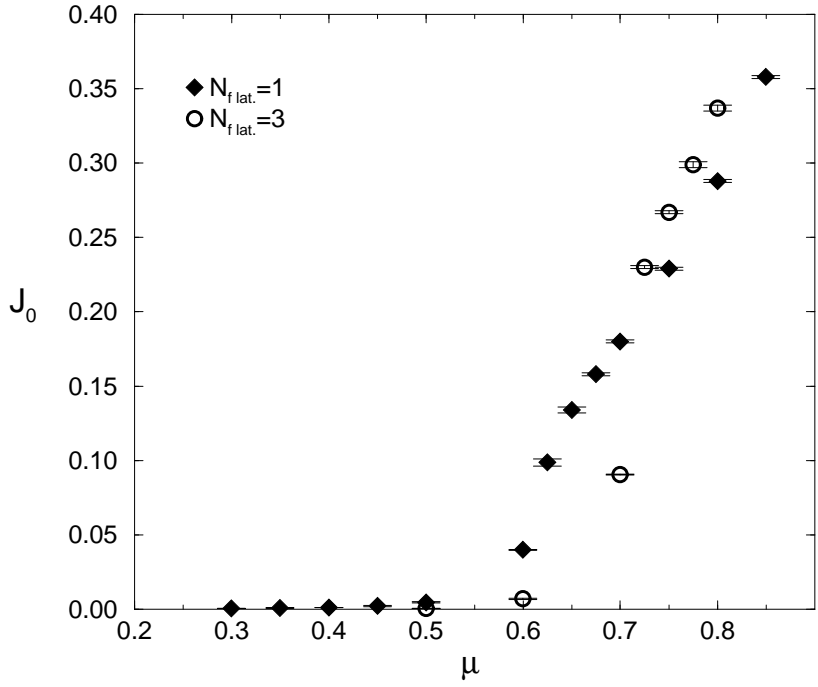,width=12cm,clip=}
\vspace{0.3cm}
\caption{Fermion number densities comparing exact hybrid Monte-Carlo 
simulations at $N=1$ with $N=3$ [6].}
\label{fig:N1_and_3}
\end{figure}

We performed standard hybrid Monte Carlo  and Glasgow Method simulations on 
$16^3$ lattices
at a four-fermi coupling of $1/g^2=0.5$ and $m=0.01$. 
To minimise the number of 
GCPF expansion coefficients we simulated the Glasgow Method with
$N=1$ flavours of staggered fermion.
(Note that the results in \cite{BIELWSP} are appropriate to $N=3$.)
In fig \ref{fig:N1_and_3} we compare number densities from standard
hybrid Monte Carlo simulations with $N=1$ and $N=3$. For $N=3$
$\mu_c=0.725(25)$ while $m_{\pi}/2=0.18(1)$ \cite{HKK3}; however for $N=1$
there is a shift such that $\mu_c\simeq 0.6$ i.e. fewer fermion flavours 
implies earlier chiral symmetry restoration. 

\begin{figure}
\epsfig{file=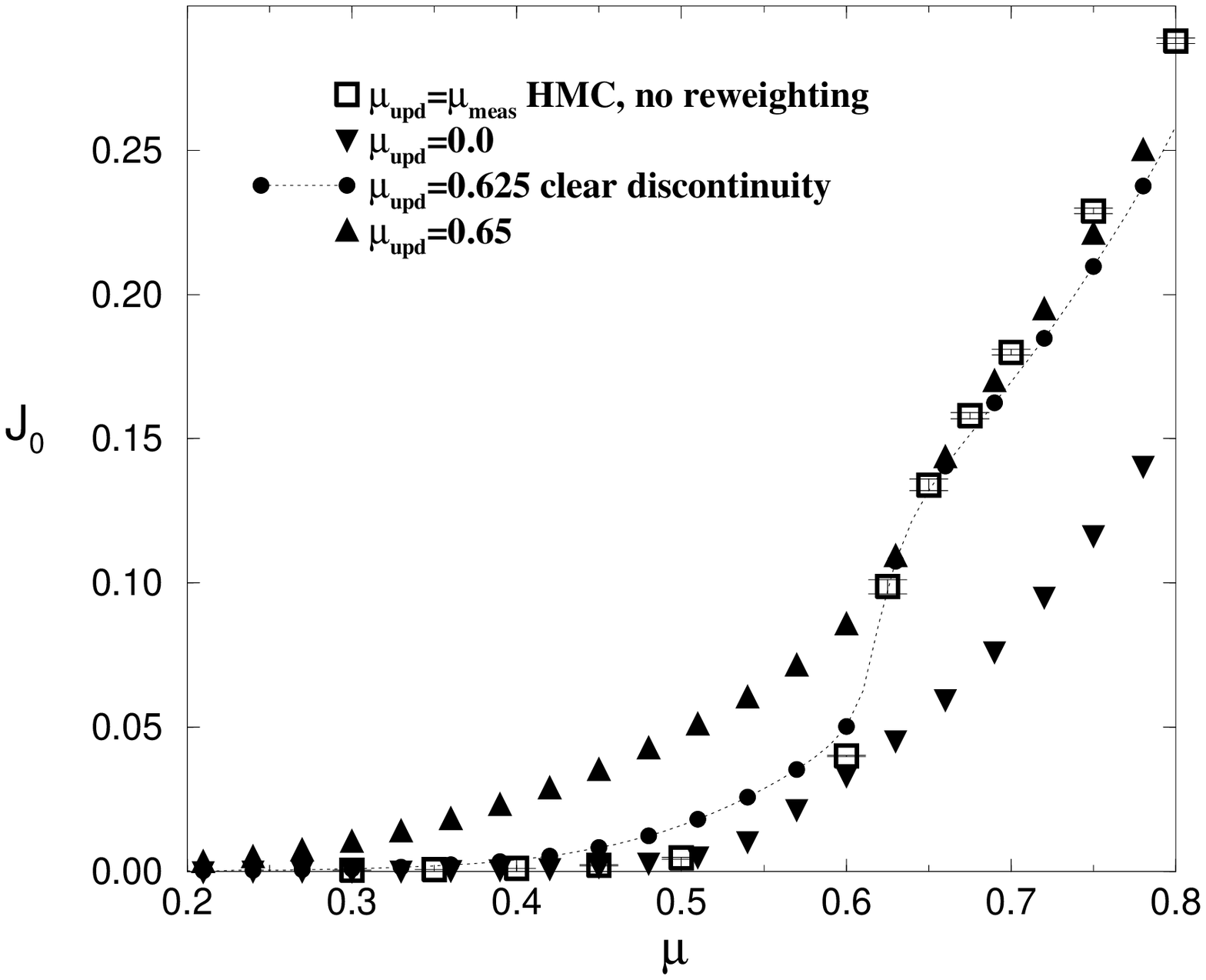,width=12cm,clip=}
\vspace{0.3cm}
\caption{Fermion number densities comparing simulations for three different 
values of $\mu_{upd}$ with the exact hybrid Monte-Carlo simulation
on a $16^3$ lattice with $1/g^{2} =0.5$, $N=1$ and $m=0.01$.}
\label{fig:all_updates}
\end{figure}

The simulation results demonstrate that the chemical potential $\mu_{upd}$ at which the
statistical ensemble is generated has a 
strong influence on the thermodynamic observables of the simulation. 
Fig. \ref{fig:all_updates} shows the number density for the standard HMC simulation and for three simulations using the Glasgow method:
one for $\mu_{upd}=0.0$ another for $\mu_{upd}=0.625(\simeq \mu_c)$ 
and finally with $\mu_{upd}=0.65(>\mu_c)$  
The discontinuity at $\mu_c$ associated with the fermion losing
dynamical mass, which is apparent in the standard hybrid Monte Carlo
 data is not consistently reproduced by the Glasgow algorithm - the results
depend strongly on the choice of $\mu_{upd}$. In fact the discontinuity
associated with the chiral transition
can be extracted from the GCPF expansion coefficients only when $\mu_{upd}\simeq\mu_{c}$.

\begin{figure}
\vspace{-0.3cm}
\centerline{\epsfig{file=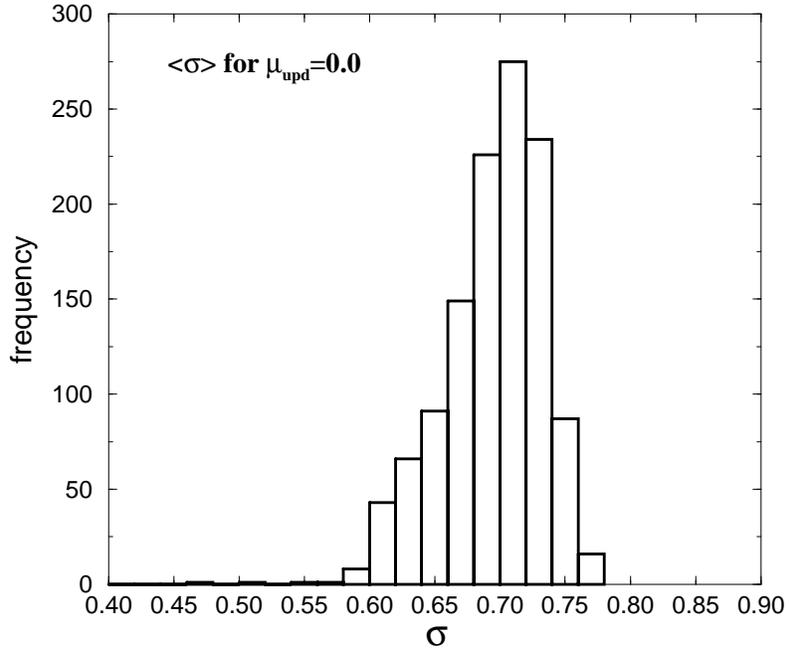,height=10cm}}
\vspace{0.1cm}
\centerline{\epsfig{file=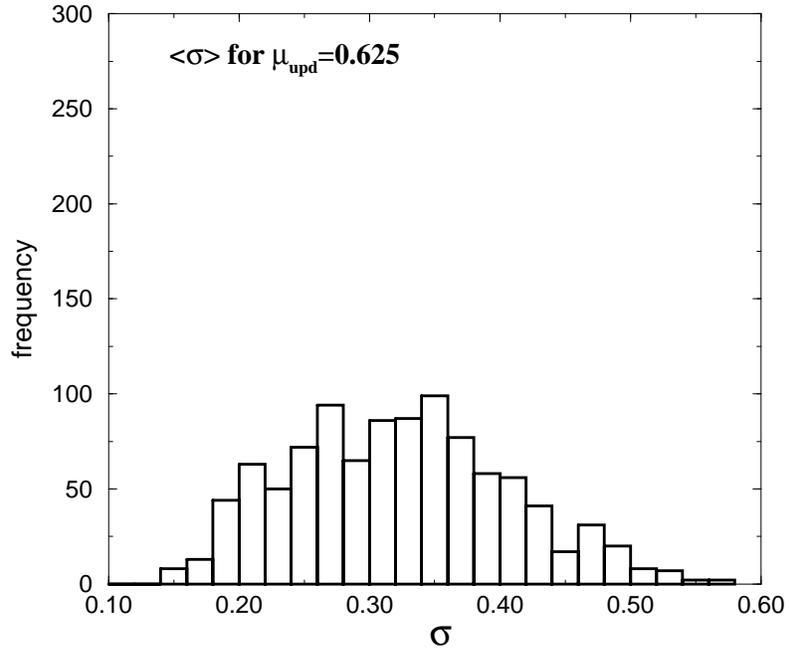,height=10cm}}
\vspace{0.3cm}
\caption{Histograms for sigma field for $\mu=0.0$ (upper) and
$\mu=0.7$ (lower) based on 1000 measurements on a $16^3$ lattice with  
$1/g^{2} =0.5$, $m=0.01$ and $N=1$. The distribution is very sharply peaked 
in the $\mu=0.0$ case and slightly broader for $\mu=0.625$ which is close
to $\mu_c$. }
\label{fig:hist_0.0_0.7}
\end{figure}

In this simple model the constraints on the effectiveness of the reweighting
must come from distribution of the $\sigma$ field.
The histograms of measurements of $\sigma$ for the two different
updates are shown in Fig. \ref{fig:hist_0.0_0.7}. For $\mu_{upd}=0$ 
the sigma field is sharply peaked about a mean value associated with a large
dynamical mass whereas for $\mu_{upd}=0.7$ the distribution
is broader although there is still no clear evidence for a 
two-state signal for a fermion with and without dynamical mass.
Despite the absence of a two-state signal in the sigma fields the
reweighting still reflects the discontinuity in the fermion number
density although this discontinuity is sharper in the standard HMC simulation.

\subsection{Free Gas behaviour}

We shall compare $J_0$ as a function of $\mu$ obtained from the Glasgow 
method with the number density of a lattice gas of free fermions with
mass $m_f=m+\langle \sigma \rangle$. Firstly we consider the simulation 
with $\mu_{upd}=0.0$ where we found $J_0$ rises smoothly from
$\mu_o$ to saturation and secondly we consider the simulation with 
$\mu_{upd}=0.625$ where we saw a discontinuity
in $J_0$ at a value of $\mu$ which was consistent with the $\mu_c$ determined by the 
standard hybrid Monte Carlo where $\mu_{upd}=\mu_{meas}$ for each value of $\mu$ simulated.

This comparison with free lattice gas number density is useful because it 
gives us a qualitative idea of whether the $J_0$ obtained for a given $\mu_{upd}$ is revealing the transition from the chirally broken phase where $m_f$
is large to the chirally symmetric phase with $m_f\simeq m$. In fact as we shall see in section V the Toy Model number density is very well approximated by 
a free lattice gas of fermions therefore any significant departure from the 
free gas behaviour for the Glasgow Method number density must be due to
the reweighting factor $R_{rw}$. 

The fermion number density for a free Fermi gas is 
obtained by summing over the finite
set of Matsubara frequencies \cite{MATSTONE84} that arises in the lattice 
thermodynamics. The
appropriate expression for the number density, $J_0$ in terms of
the momentum sums for a 3-dimensional free lattice 
gas is :-
\begin{equation}
{J_0}_{free}=2i \int_{-\pi}^{\pi}\frac{d^{3}p}{{2\pi}^4}\, \frac {\sin(p_t + i\mu)\cos(p_t +i\mu)}
{\sum_{i=1}^{2}\sin^{2}p_{i} + \sin(p_{t} + i\mu) + m_{f}^2}
\end{equation}
where $p_t$ is associated with the temporal direction.

\begin{figure}
\vspace{-0.3cm}
\centerline{\epsfig{file=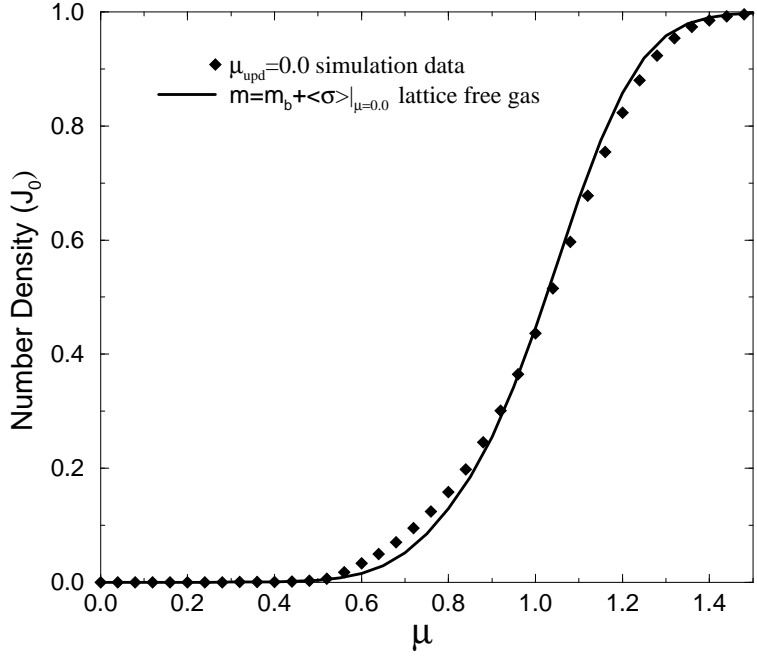,height=10cm}}
\vspace{0.1cm}
\centerline{\epsfig{file=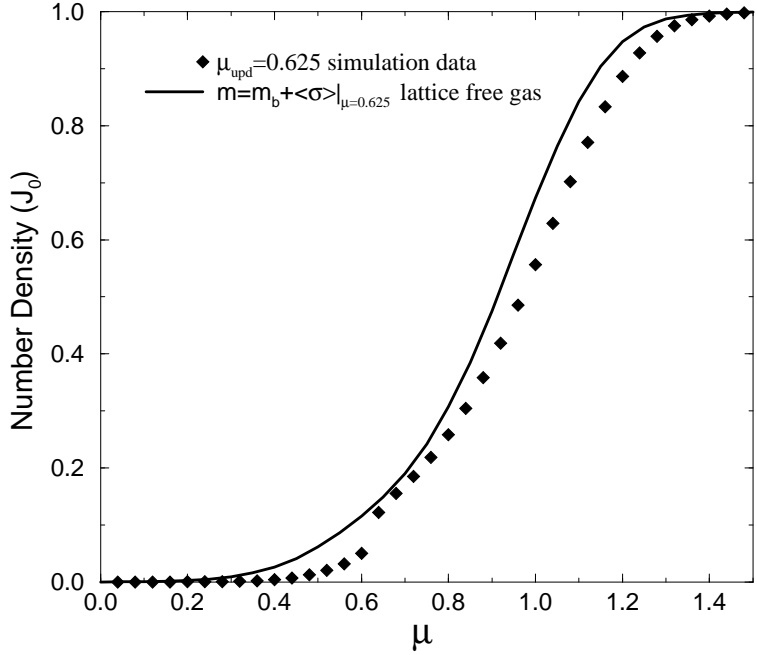,height=10cm}}
\vspace{0.1cm}
\caption{Comparison of number densities ($J_0$) for free lattice gases of fermions
of specified mass, m  with the corresponding $J_0$ from Gross-Neveu simulations
using the Glasgow method for $\mu_{upd}=0.0$ (upper) and $\mu_{upd}=0.625$ (lower) on a $16^3$ lattice with $1/g^{2} =0.5$, $m=0.01$ and $N=1$.}
\label{fig:fgas}
\end{figure}

Recall that $\langle\sigma\rangle$ is a measure of the physical fermion mass.
The results of our $J_0$ comparison are plotted in Fig. \ref{fig:fgas}. 
In our GN simulations using the Glasgow method for $\mu_{upd}=0.0$ the fermion 
number density was consistent with a gas of free fermions with a dynamical mass
$m_{f}=m+\langle \sigma \rangle|_{\mu=0}$.  
As we shall describe in section V, this is the same behaviour we observed 
for the Toy model.
However for $\mu_{upd}=0.625$ there is a clear indication of a 
{\it discontinuity} at $\mu_c$ and for $\mu>\mu_c$ the results are 
consistent with a free gas of fermions with dynamical mass 
$m_b+\langle\sigma\rangle|_{\mu=0.625}$. 
We found that for $\mu_{upd}=0.65$  i.e. as the chemical potential in
the functional measure increased beyond $\mu_c$ the discontinuity disappeared
again. This suggests that within the limits of the statistics of a typical lattice 
simulation, the inclusion of the reweighting term $R_{rw}$ in the expression
for the GCPF is insufficient to project onto the ensemble appropriate
to $\mu_{meas}$ except when $\mu_{meas}\simeq\mu_{upd}$.

\subsection{Lee-Yang Zeros}

\begin{figure}
\vspace{-0.3cm}
\centerline{\epsfig{file=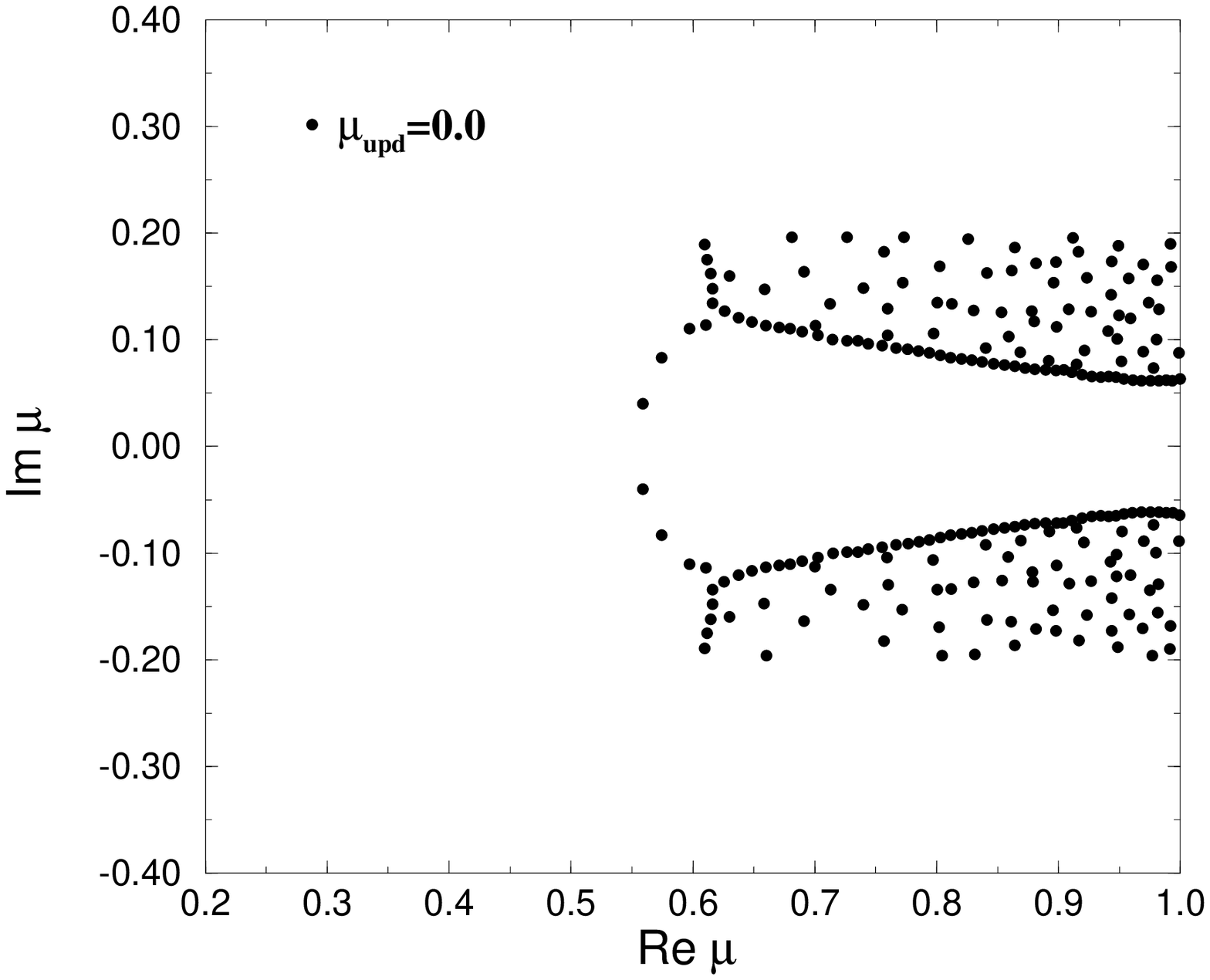,width=6.5cm}}
\vspace{0.5cm}
\centerline{\epsfig{file=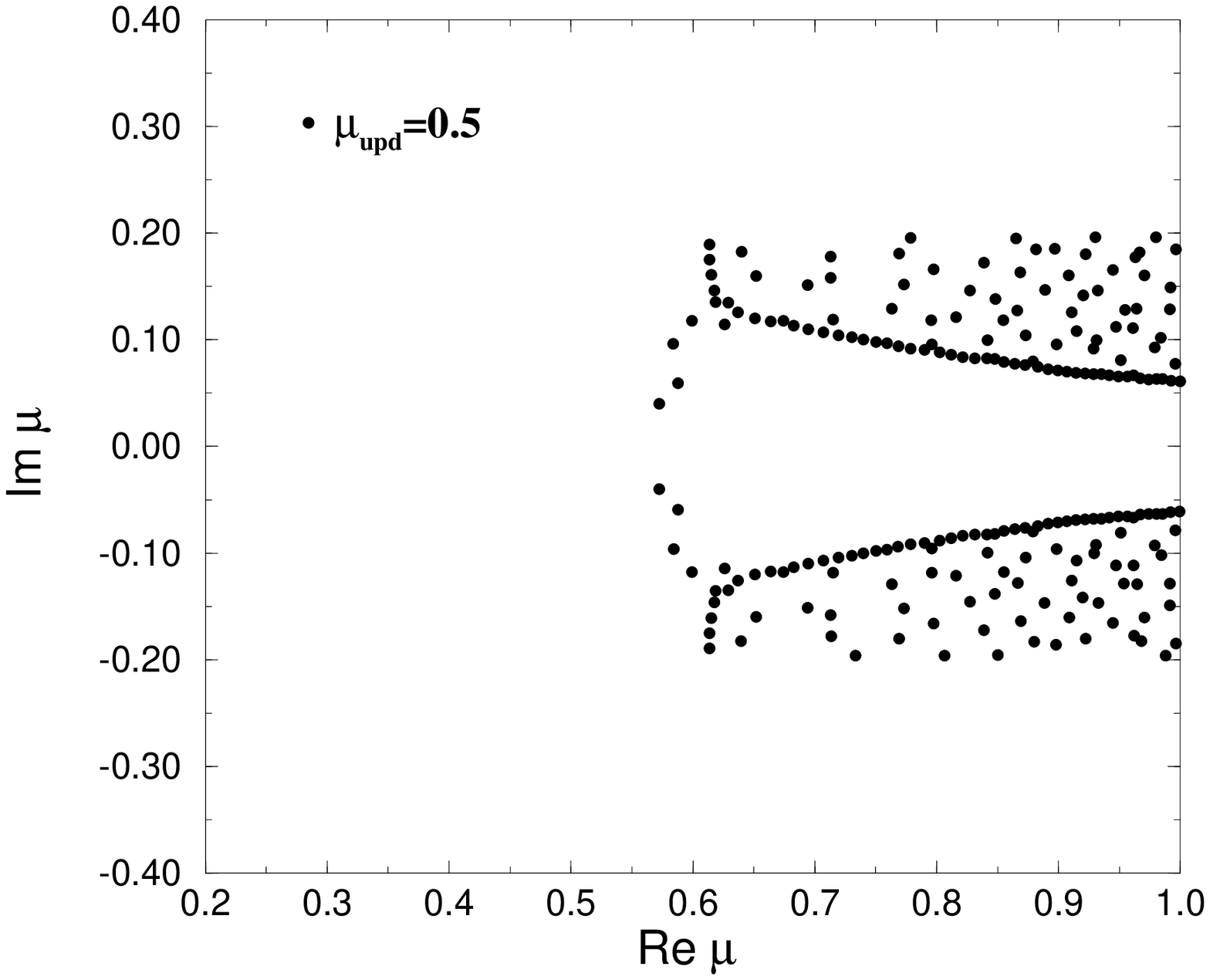,width=6.5cm}}
\vspace{0.5cm}
\centerline{\epsfig{file=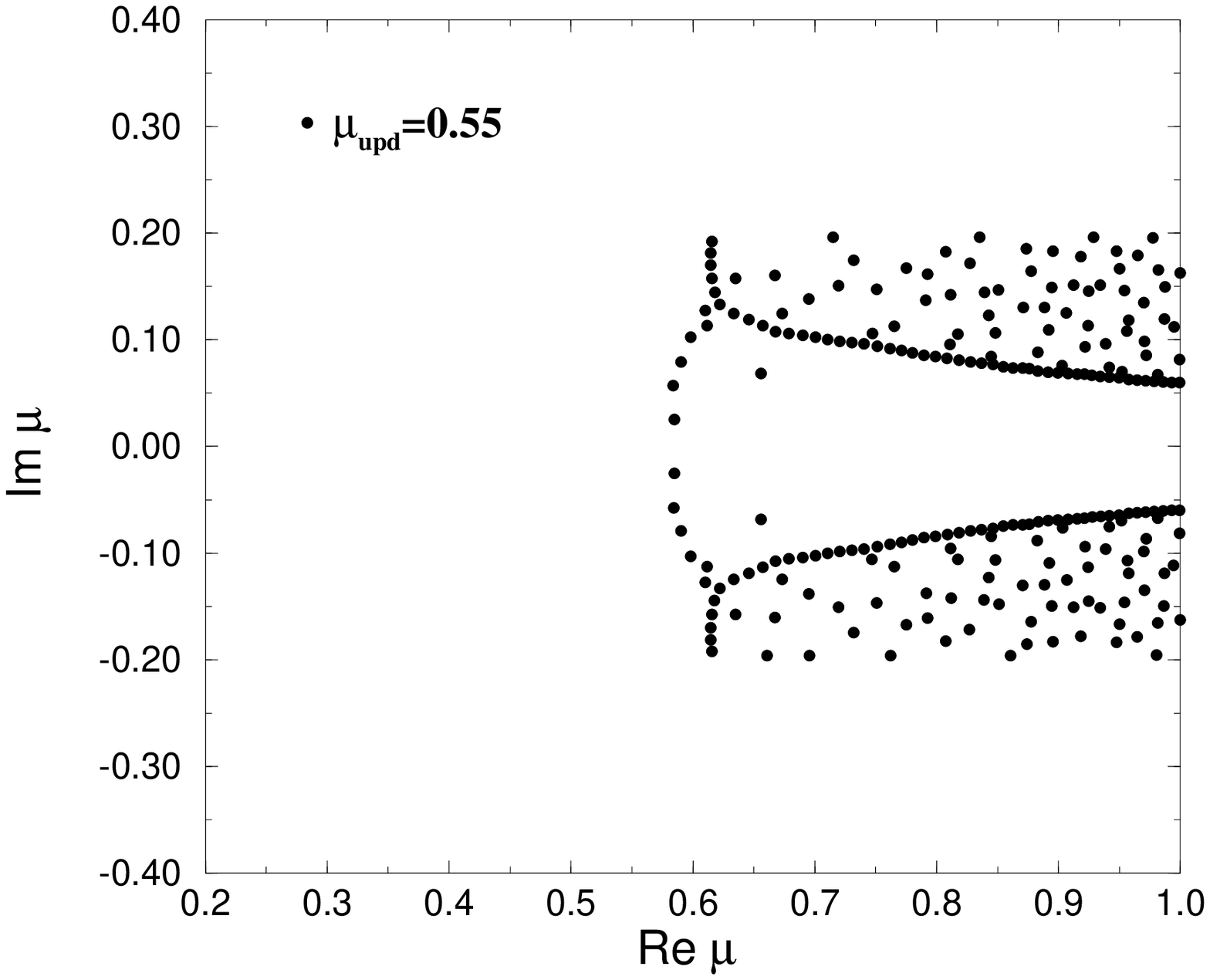,width=6.5cm}}
\vspace{0.5cm}
\centerline{\epsfig{file=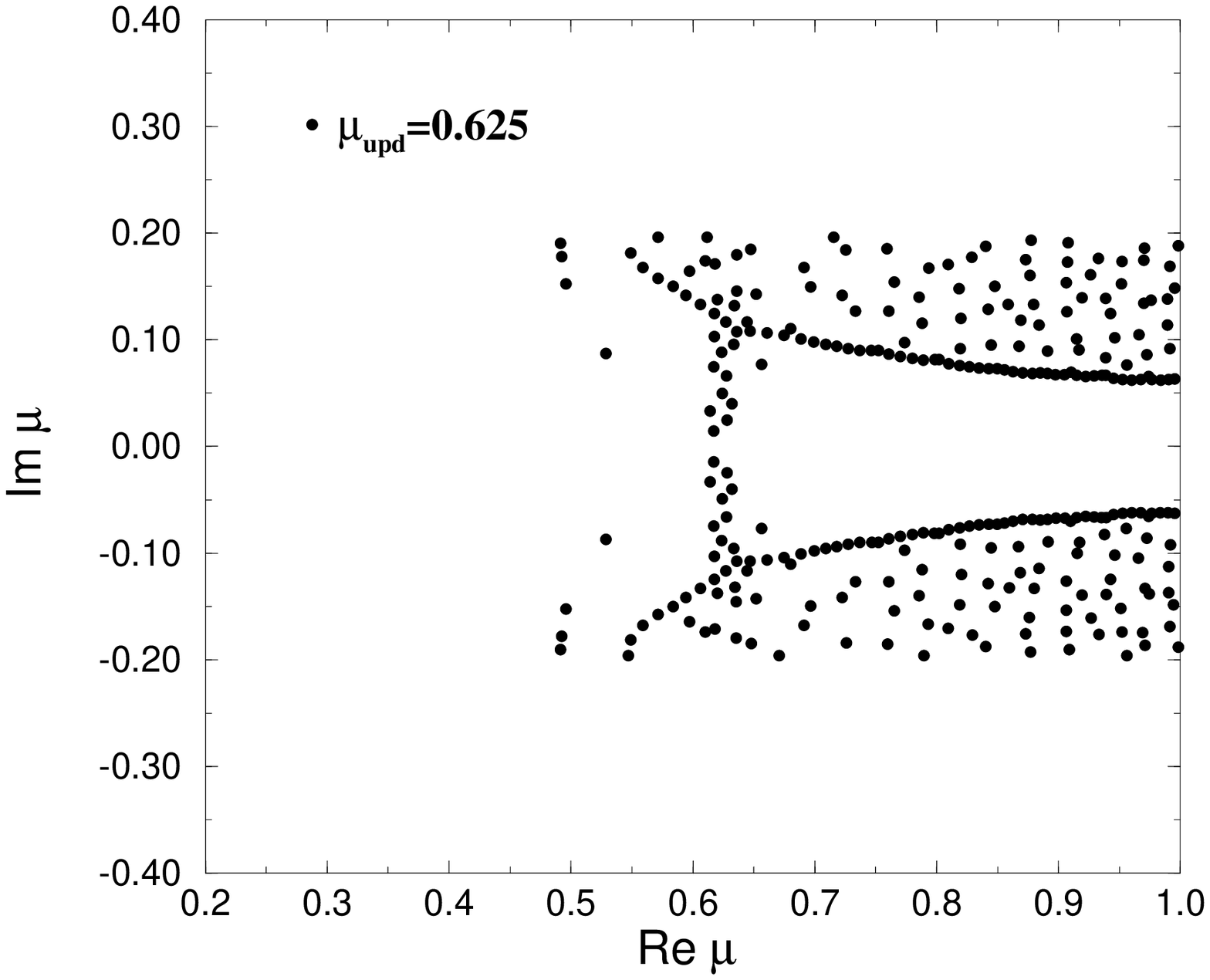,width=6.5cm}}
\vspace{0.5cm}
\centerline{\epsfig{file=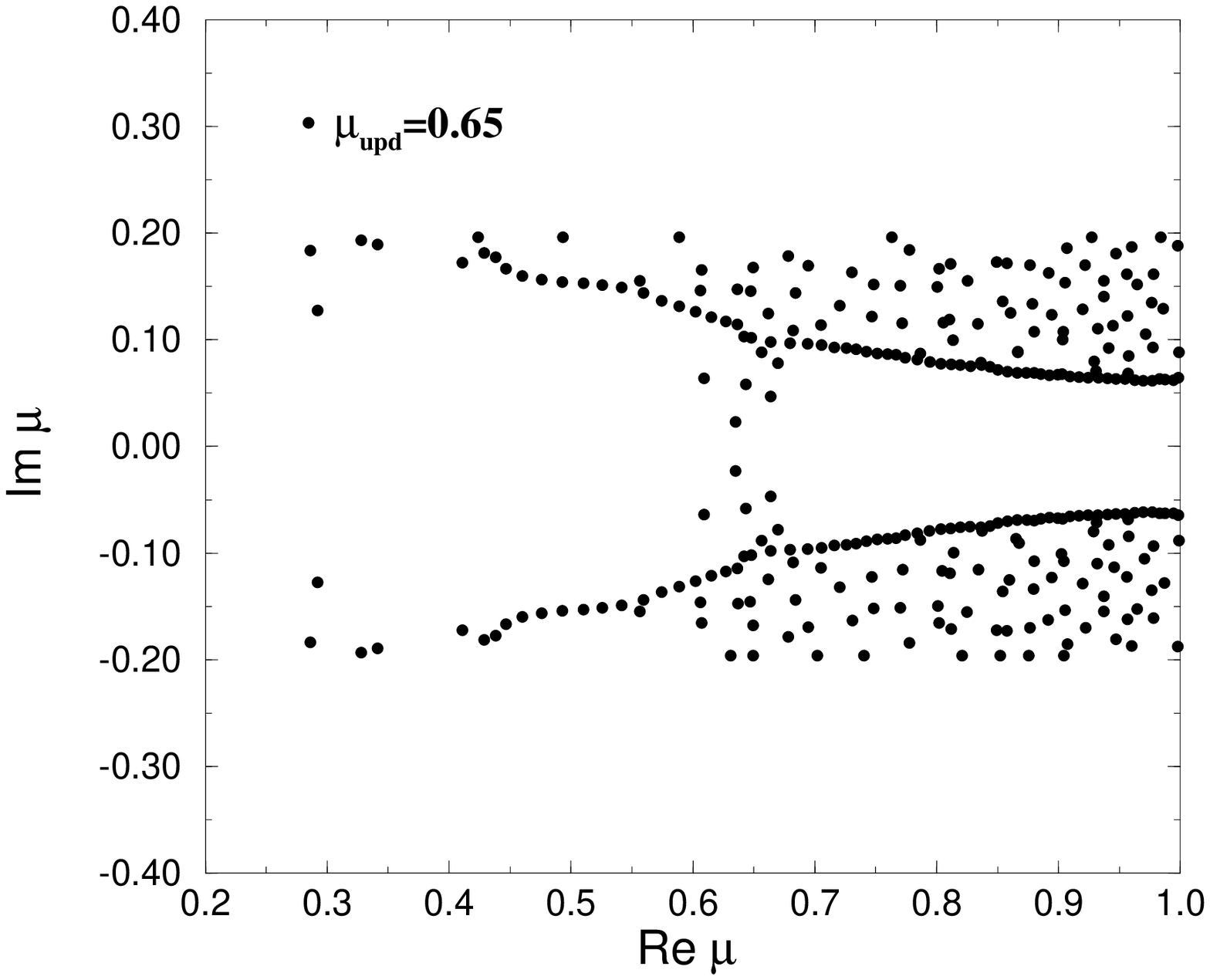,width=6.5cm}}
\vspace{0.5cm}
\centerline{\epsfig{file=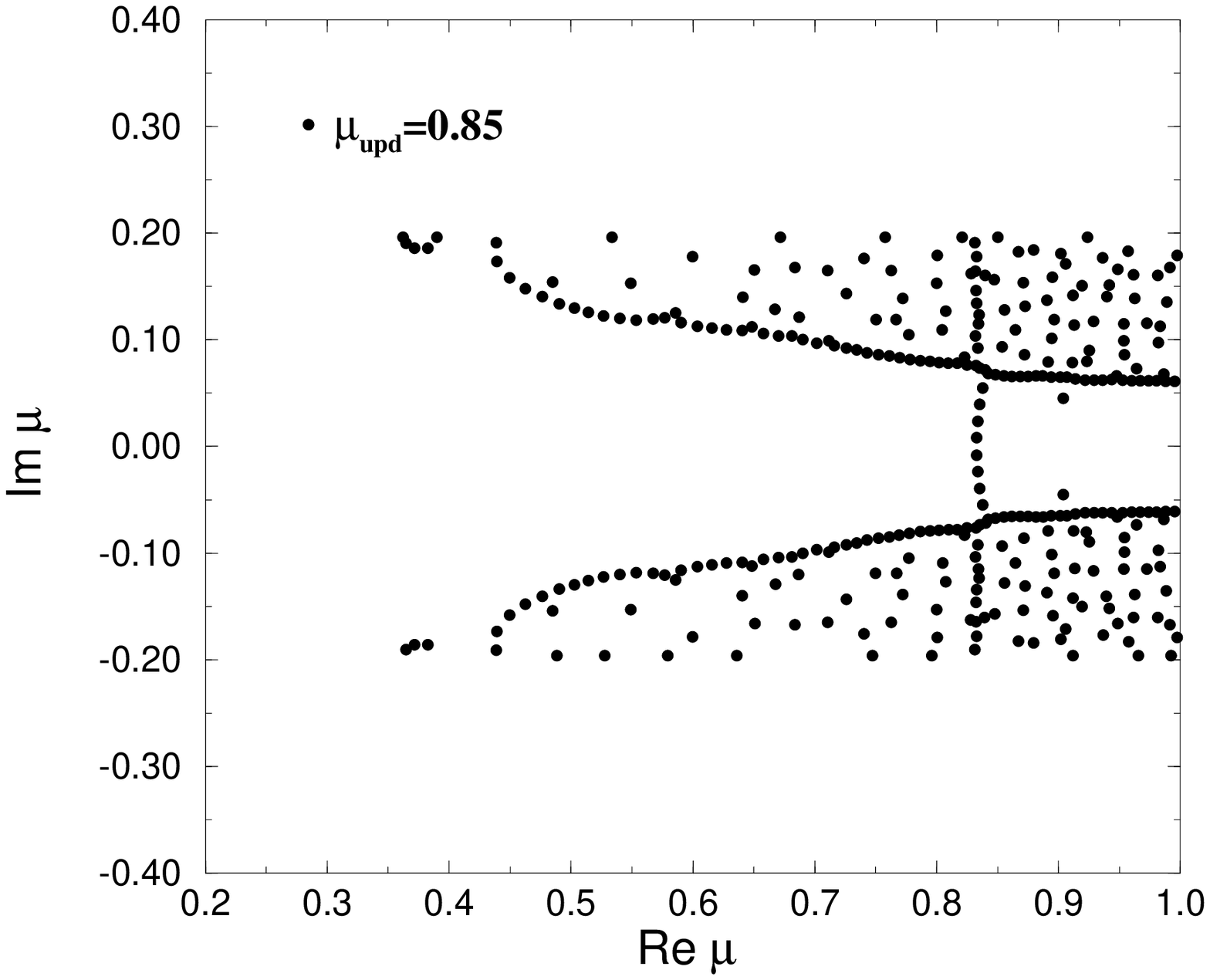,width=6.5cm}}
\vspace{0.5cm}
\caption{Partition function zeros for six values of $\mu_{upd}$.
 Simulations on $16^3$ lattice with $N=1$, $1/g^{2} =0.5$ and $m=0.01$.}
\label{fig:zeros}
\end{figure}

We have seen that the Glasgow Method fermion number density only gives a reliable estimate for $\mu_c$ from the number density when $\mu_{upd}\simeq \mu_c$: 
however we also expect a signal 
for criticality from the partition function zeros.
According to the theorems of Lee and Yang \cite{LEEYANG}, the phase 
transitions of a system are controlled by the distribution of roots of 
the GCPF. A phase transition occurs whenever a root approaches the real 
axis in the infinite volume limit. In practical lattice simulations 
we are certainly not close to the thermodynamic limit, however
as the lattice volume is increased we expect the zero with the
smallest imaginary part to approach the real axis, converging to
the critical value of the appropriate parameter. The Lee-Yang zeros 
in the complex $\mu$ plane are the zeros of 
eqn. (\ref{eqn:GCPF}) and their distribution should reflect $\mu_c$. 
The zeros in the complex $\mu$ plane for the 
$\mu_{upd}=0.0,0.5,0.55,0.625,0.65,0.85$  are plotted in 
Fig.\,\ref{fig:zeros}. The first point to note is that the zeros form
a distinctive and elaborate pattern. Even for this simple model it 
would be difficult to predict analytically the distribution of zeros
however some very clear features emerge when we compare the distributions 
appropriate to different $\mu_{upd}$. For $\mu_{upd}<\mu_c$ i.e.
$\mu_{upd}=0.0,0.5,0.55$ there is an arc of zeros which intersects
the real axis at $\mu\simeq0.6\simeq\mu_c$. For these 3 values of
$\mu_{upd}$ we saw no evidence of a discontinuity in $J_0$ at
$\mu=0.6$. Now at $\mu_{upd}=0.625$ there is clear change in the pattern
of zeros. What was an arc of zeros at $\mu\simeq0.6$ now becomes a broader 
vertical line and two additional lines of zeros sit outside the main body
of the distribution at $\mu\simeq0.5$. Recall that for $\mu_{upd}=0.625$ 
we did see a discontinuity at $\mu_c$ in $J_0$. As we increase $\mu_{upd}$
beyond $\mu_c$ i.e. for $\mu_{upd}=0.65,0.85$ the vertical line of zeros 
migrates to higher $\mu$ and the main zeros distribution extends further into
the low $\mu$ region of the complex $\mu$ plane. For $\mu_{upd}=0.85$ the 
vertical line of zeros sits at $\mu\simeq 0.85$ i.e. far from $\mu_c$.

\section{Toy Model versus standard hybrid Monte Carlo results}

In this section we  discuss an approximate model, which we will
refer to as the `Toy' model, which will turn out to give useful insight 
into the results of Glasgow reweighting applied to the GN model. In the 
Toy Model the chemical potential used in the measure $\mu_{meas}$
can be difefrent from that used in the update $\mu_{upd}$.
It can be seen as the equivalent of the quenched 
approximation when $\mu_{upd} = 0$, and is the exact model 
when $\mu_{meas} = \mu_{upd}$. 
It is essentially the Glasgow Method with $R_{rw}\equiv1$.
When $\mu_{upd}=0$ and the statistics is low, the Glasgow
method results coincide with thos of the Toy  model, and 
should approach the exact results in the large statistics limit \cite{SC}.

Let us begin by  reviewing what is known about the GN model in the mean
field approximation.

In the mean field approximation, which turns out to be the 
leading order of an expansion of powers of $1/N_f$,
we solve for the expectation value of the 
auxiliary scalar $\langle\sigma\rangle$ self-consistently using the gap 
equation, which reads (with bare mass $m$ set to zero):
\begin{equation}
\langle\sigma\rangle = -g^2 \langle\bar \psi \psi\rangle = 
{g^2\over V} \mbox{tr} S_F(\langle\sigma\rangle)=\int_p\mbox{tr}
g^2{1\over{ip{\!\!\!/}\,+\langle\sigma\rangle}}.
\label{for:gap}
\end{equation}
Eqn. (\ref{for:gap}) can be solved using a lattice regularisation \cite{HKK_AP}
and has a non-trivial solution $\langle\sigma\rangle\not=0$ for 
$g^2>g^2_c\simeq1.0$. The dynamical fermion mass 
$m_f=\langle\sigma\rangle$ then 
defines the model's physical scale,
and there is a continuum limit corresponding
to a continuous chiral symmetry restoring phase transition as 
$\langle\sigma\rangle\searrow0$, $g^2\searrow
g_c^2$. 

The gap equation can also be generalised to $\mu\not=0$:
\begin{equation}
\langle\sigma\rangle(\mu) = -g^2 \langle\bar \psi \psi\rangle(\mu) = 
{g^2\over V} \mbox{tr} S_F(\mu,\langle\sigma\rangle(\mu));
\label{for:gapmu}
\end{equation}
once again the solution can be found, either in the continuum \cite{RWP},
or on a lattice \cite{HKK4}\cite{HKK3}.
The salient feature of the solution is that $\langle\sigma\rangle(\mu)$ 
remains unchanged from its zero-density value $\langle\sigma\rangle(0)$
as $\mu$ is increased, up to a critical value
\begin{equation}
\mu_c = \langle \sigma \rangle(0),
\end{equation}
whereupon it falls immediately to zero, signifying that chiral symmetry 
is restored via a first order transition. Introduction of
a small bare fermion mass $m$ causes a slight softening of the transition.
The hybrid Monte Carlo simulation results of 
refs. \cite{HKK4}\cite{HKK3} confirm that this picture is qualitatively 
correct.

Now consider
a 3d GN model in which a value of the chemical potential $\mu_{upd}$ is used 
to generate the ensemble, and a value $\mu_{meas}$ to measure
expectation values, with $\mu_{upd}\not=\mu_{meas}$ in general.
We will refer to this as the `{\sl Toy model\/}'
Notice that $\mu_{upd} = \mu_{meas}$ defines the standard hybrid Monte Carlo
 calculation
whereas $\mu_{upd} = 0$ is the equivalent of a quenched study (we could indeed 
refer to the case $\mu_{upd}\not=\mu_{meas}$ as a `generalised quenched
approximation'). 

In a mean field treatment of the Toy model 
the $\sigma$ field  is constrained to
its $\mu = \mu_{upd}$ value, so therefore we simply substitute 
$\langle\sigma\rangle = \langle\sigma\rangle(\mu_{upd})$ 
in the fermion propagator, and calculate
$\langle\bar\psi\psi\rangle$. Therefore
the Toy counterpart of eqn.(\ref{for:gapmu})  is
\begin{equation}
\langle\sigma\rangle(\mu_{upd}) \not= -g^2 <\bar \psi \psi>(\mu_{meas}) = 
{g^2\over V}\mbox{tr} S_F (\mu_{meas}, \langle\sigma\rangle(\mu_{upd}))
\label{eq:mftoy}
\end{equation}
%

In the mean field approximation, the above considerations suggest
that all Toy models are equivalent to free fermion models with 
$m_f=\langle\sigma(\mu_{upd})\rangle$. This
is confirmed by the numerical simulations described below. It is clear that 
Glasgow method reweighting cannot work in the GN model because 
the ensemble generated with $\mu_{upd} < \mu_c$ has zero overlap
with that generated using
$\mu_{upd} > \mu_c$.
This means that all Toy models with $\mu_{upd} < \mu_c$ should be equivalent,
and should approximate the standard HMC results for $\mu_{meas}<\mu_c$.
However, for $\mu_{meas}\geq\mu_c$ the propagator $S_F$ remains
that of a free fermion with $m_f=\langle\sigma\rangle(\mu_{upd})$ i.e. the fermion appears to retain its dynamical mass
therefore we cannot expect a chiral symmetry restoring transition in the 
Toy model.
Consider next the Toy model with $\mu_{upd} > \mu_c$.
The propagator $S_F$ now always describes a massless fermion, therefore in this case we expect 
the Toy model to reproduce the exact results only for $\mu_{meas}>\mu_c$.

\begin{figure}
\centerline{\epsfig{file=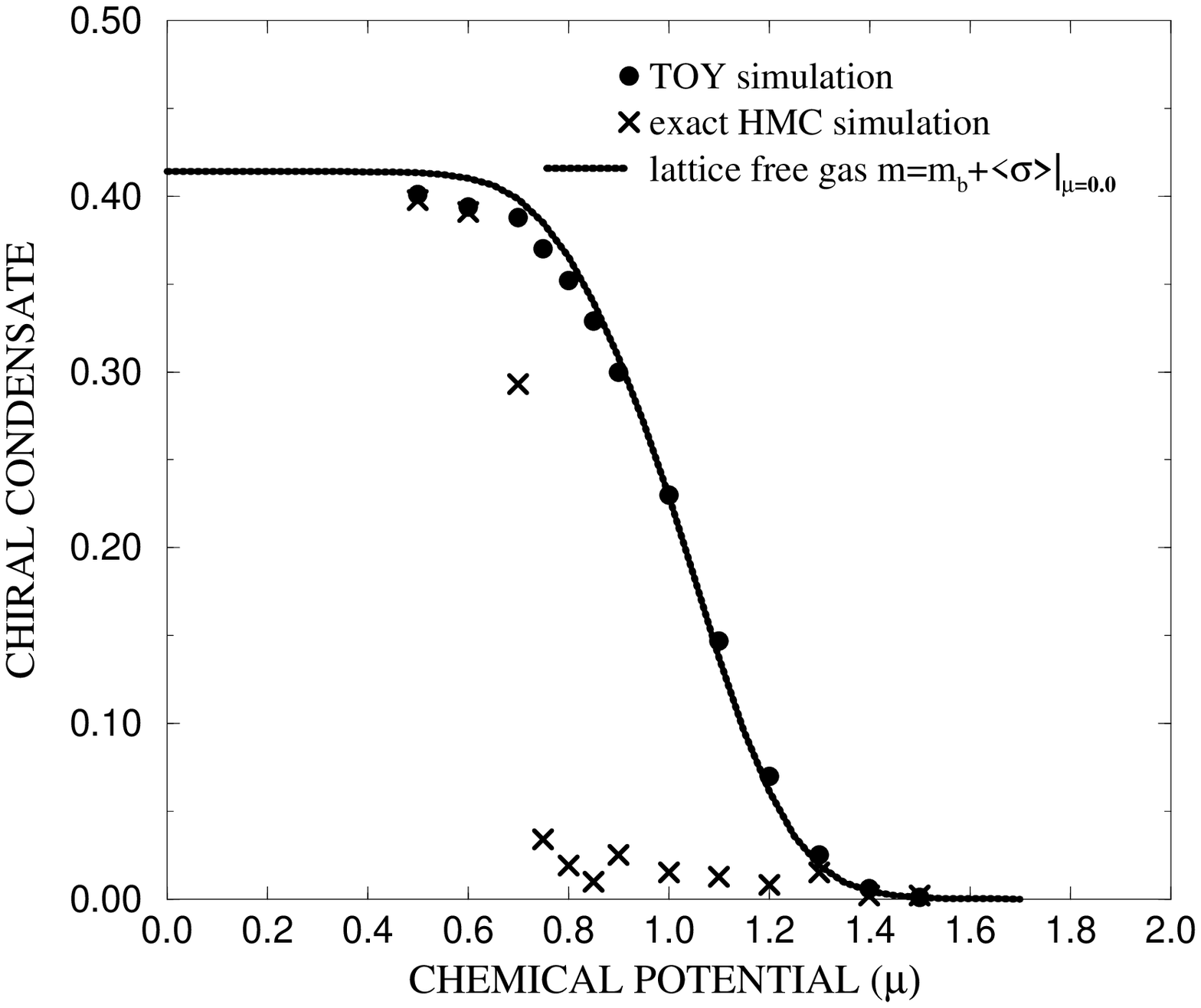,width=10cm}}
\centerline{\epsfig{file=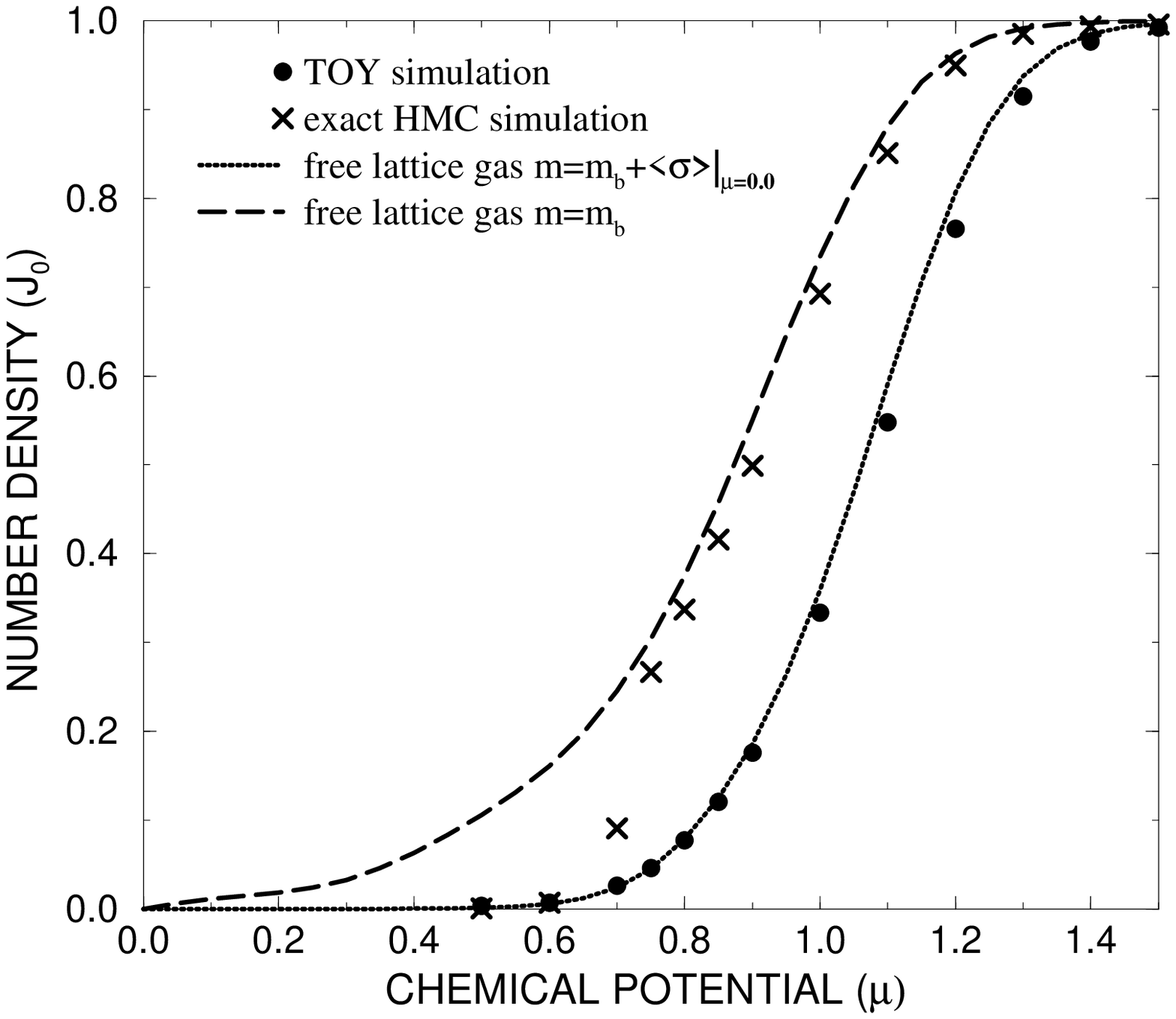,width=10cm}}
\caption{$<\bar\psi\psi>$ (upper) and $J_0$ in the
full and Toy GN model at $\beta = 0.5$, $m_q = 0.01$, 
$L=16$, $N = 3$.
The data for the full model $\mu < 0.9$ are from [6]}
\label{fig:pbp_num_gn}
\end{figure}

It is also possible, and instructive, to test these ideas once
quantum fluctuations of $\sigma$ are included. Accordingly we have performed
simulations of the Toy model for $\mu_{upd}=0$, $\mu_{meas}$ arbitrary, 
using the hybrid Monte Carlo algorithm.
Fig. \ref{fig:pbp_num_gn} compares the results of these simulations
with the exact simulation results \cite{HKK3}, obtained with the same
parameters (coupling $1/g^2=0.5$, bare mass $m=0.01$, 
lattice size $L=16$, $N=3$ corresponding to $N_f=12$ continuum flavours), 
and the mean
field predictions of (\ref{eq:mftoy}).
Both the condensate $\langle\bar\psi\psi\rangle$ and number density
$\langle J_0\rangle$ are plotted as functions of $\mu_{meas}$.
We see that the Toy model with
$\mu_{upd} = 0$ is well described by a massive free field with
$m_f = \langle\sigma\rangle(0)$ for $\mu_{meas}<\mu_c\simeq0.725$, and that 
exact and Toy results agree well in this range; 
for $\mu_{meas}>\mu_c$, however, while the
exact results for $\langle\bar\psi\psi\rangle$ fall steeply,
the Toy results exhibit a smooth crossover, approaching zero only
for $\mu_{meas}\simeq1.4$. 
For $\mu_{meas}>\mu_c$ the full model is better described by a massless
free field, corresponding to $\mu_{upd}>\mu_c$,
$\langle\sigma\rangle(\mu_{upd})=0$,
indicating the restoration of chiral symmetry.
In all cases the mean field prediction (\ref{eq:mftoy}) reproduces the
Toy simulation results well, indicating that quantum fluctuations are 
small due to the relatively large value of $N_f$.

We deduce that
$\mu_c$ is not observable in
the Toy model, since there is no chiral symmetry restoration. 
Note however that the value $\mu_o$ at which the onset of
$\langle J_0\rangle$ occurs is the same for both Toy and 
exact models, reflecting the fact that the onset on a finite lattice (and
perhaps even in the thermodynamic limit) is
not associated with the chiral transition. 

We emphasize that this simple analysis works well because 
at $1/g^2 = 0.5$ and $N_f=12$ the fluctuations
in $\sigma$ are highly suppressed. We verified that the  microscopic 
distribution
of the $\sigma$ field is very sharply peaked about its constant 
value. This condition serves to makes the model
well approximated by the appropriate free fermion field in each phase.
Because the two phases for $\mu < \mu_c$ and $\mu >\mu_c$
are clearly
different, we anticipate
that Glasgow reweighting using $\mu_{upd}=0$
will not be able to capture the chiral
phase transition because the overlap between $\sigma$ distributuions
from each phase is likely to be very small. 
The overlap would be exactly zero in the 
infinite $N_f$ limit where fluctuations 
are completely suppressed. 
It would be interesting to repeat this exercise closer to
the critical point $g^2_c \simeq 1$ although at this coupling
the chiral transition is more difficult to identify.

\section{Pseudoscalar Mass and Realisation of the Goldstone Mechanism}

Although we have shown that the Glasgow method and the Toy model 
simulations differ in their quantitative predictions for the
number density $J_0(\mu)$ from the standard hybrid Monte Carlo 
simulations we observe that the {\sl onset}, $\mu_{o}$ in $J_0$ (ie. the value
of $\mu$ where $J_0$ starts to increase from zero which is usually
interpreted as the point of chiral symmetry restoration) {\sl is} correctly 
predicted by both Toy model and the Glasgow method. This is true even for 
simulations with modest statistics.

 An important result reported in \cite{HKK3} was the numerical proof  that $\mu_{o}\gg m_{\pi}/2$
in standard hybrid Monte Carlo simulations of the GN model. Thus we had an example of a theory which could be simulated at $\mu\neq 0$ with dynamical fermions and where there were no pathologies associated with the fact that there was
a Goldstone pion in the spectrum. In the Gross-Neveu theory even in the Toy model simulations (which are partially quenched) $\mu_{o} \gg m_{\pi}/2$ in 
agreement with the full theory. Why then are there pathologies associated
with $m_{\pi}$ in QCD? We set out to gain insight into this fundamental
difference between the two lattice models.

We shall discover that the onset 
in $J_0$ always occurs at  $\mu_{o}=m_{PS}/2$, where $m_{PS}$ is the mass of
the pseudoscalar state measured via $GG^{\dagger}$ where $G$($G^{\dagger}$)
is the fermion (anti-fermion) propagator .
In QCD $m_{PS}$ {\bf is} the Goldstone pion, however in four fermion models
the Goldstone mechanism is realized by the auxiliary field
$\vec \pi$ and  $m_{PS}$ is a much heavier state 
which appears to be close to twice the dynamical fermion mass.

Let us first consider the realisation of the Goldstone mechanism in the Gross-Neveu model. In particular we measure and compare the connected and disconnected contributions to the Goldstone pion. If the pion has a dominant connected contribution we expect a pole in the pseudoscalar propagator 
 corresponding to a light particle with mass$\propto\sqrt{m}$. If, on the other hand, the pion has a dominant disconnected contribution the mass scale corresponding to the pole in the pseudoscalar propagator will be 
$\simeq 2m_f(\mu=0)>>m_{\pi}$.   

Consider the lattice Ward identity for the chiral condensate: 

\begin{eqnarray}
\sum_{y} \langle {\bar{\psi}\gamma_{5}\psi(y)\bar{\psi}
\gamma_{5}\psi(x)}\rangle 
&=&\sum_{y}\langle{tr(G_{-\mu}^{\dagger}(x,y)G_{+\mu}(x,y))}\rangle-
\langle {(tr\gamma_{5}G(x,x))(tr\gamma_{5}G(y,y))}\rangle  \nonumber \\
&=& -{\frac{\langle {\bar{\psi}(x)\psi(x)}\rangle}{m_q}}
\label{eqn:ward}
\end{eqnarray}

The pion susceptibility Eqn.(\ref{eqn:ward})
has contributions from a connected channel (1st term) and a disconnected 
channel (2nd term).

Consider the Dirac kinetic operator $D{\!\!\!\! /}\,$ for staggered fermions. 
At $\mu=0$ the 
relation $D{\!\!\!\! /}\,^{\dagger}=-D{\!\!\!\! /}\,$ holds.
For $\mu\neq 0$  the fact that $e^{\mu}$ ($e^{-\mu}$) multiplies the 
forward (backward) timelike gauge
links in the lattice action means that ${D\!\!\!\! /}\,^{\dagger}
\neq-D{\!\!\!\! /}\,$, because in the matrix
$D{\!\!\!\! /}\,^{\dagger}$, 
$e^{-\mu}$ ($e^{\mu}$) now multiplies the forward (backward)
links. Consequently the propagators $G$ (with $G=M^{-1}$) and 
$G^{\dagger}$ are no longer trivially related as they are at $\mu=0$.
Thus the pseudoscalar propagator, $G_{ps}$, is defined by

\begin{equation}
G_{ps}(t)=\sum_{\vec{x}} G_{+\mu}(\vec{x},t)G_{-\mu}^{\dagger}(\vec{x},t) 
\end{equation}

Let $\lambda_{ps}$ be the smallest eigenvalue of the 
Gibbs propagator matrix $P$ (eqn. \ref{eqn:propmat}). It can be shown that $\lambda_{ps}$ is
associated with the mass pole in $G_{ps}$ and that when we 
consider a single lattice configuration $\lambda_{ps}$ directly corresponds
to a Lee-Yang zero and therefore induces a singularity in $J_0$ i.e. triggers
the rise from zero of the fermion number density. To see this 
observe that $\det M$ can be simply expressed in terms of the eigenvalues of $P$
and $Z$ can be similarly expressed in terms of its zeros $\alpha_i$ in the $e^{\mu}$ plane:
\begin{equation}
\det M \propto \prod_{i}(e^{\mu}-\lambda_i) \;\;\;;\;\;\;
Z \propto \prod_{i}(e^{\mu}-\alpha_i).
\end{equation}

Since $Z = \langle \det (M)\rangle$ we see that on a single configuration 
$\alpha_i=\lambda_i$. Note however that the ensemble
averaged $\alpha_i$'s are not in general the same as the ensemble 
averages of the $\lambda_i$'s.  

The number density on a single configuration 
$J_0^i\sim {{\partial \ln \det M}/{\partial \mu}}$ whereas the ensemble
average is given by $J_0\sim {{\partial \ln \langle\det M\rangle}/{\partial \mu}}$. Clearly the chemical potential where the number density begins to rise from zero on a single configuration will
be determined by the numerical value of $\lambda_{ps}$ 
(i.e. $\mu_{o}=\lambda_{ps}$). 

It has been proved by Gibbs \cite{GIBBS2} that if we assume that the
pion has a dominant connected contribution it follows that  
$\lambda_{ps}\sim e^{-\frac{m_{\pi}}{2}}$. In this case we can only  
envisage achieving a physically meaningful result
($\mu_o>\frac{m_{\pi}}{2}$) if  $\alpha_i\neq\lambda_i$. 
However in the exceptional circumstance where the 
disconnected contribution to the pion is dominant we expect
that $\lambda_{ps}\sim e^{-m_{f}}$ so that configuration by configuration
$\mu_{o}=m_f>\frac{m_{\pi}}{2}$. 
In this section we shall prove that in the 3d GN $U(1)$ model the Goldstone pole forms in the disconnected channel therefore the state described by $G_{ps}$
no longer corresponds to the Goldstone pion. Instead we find:
\begin{equation}
G_{ps} \simeq e^{-2m_f(\mu=0)t}
\label{GN_pion}
\end{equation}
Since $|\lambda_{ps}|$  will now correspond to the dynamical fermion
mass $m_f$ rather than $m_{\pi}/2$ we have no reason 
to expect an early onset in the Gross-Neveu model despite the existence of a light Goldstone pion in the spectrum.

\begin{figure}
\vspace{0.2cm}
\epsfig{file=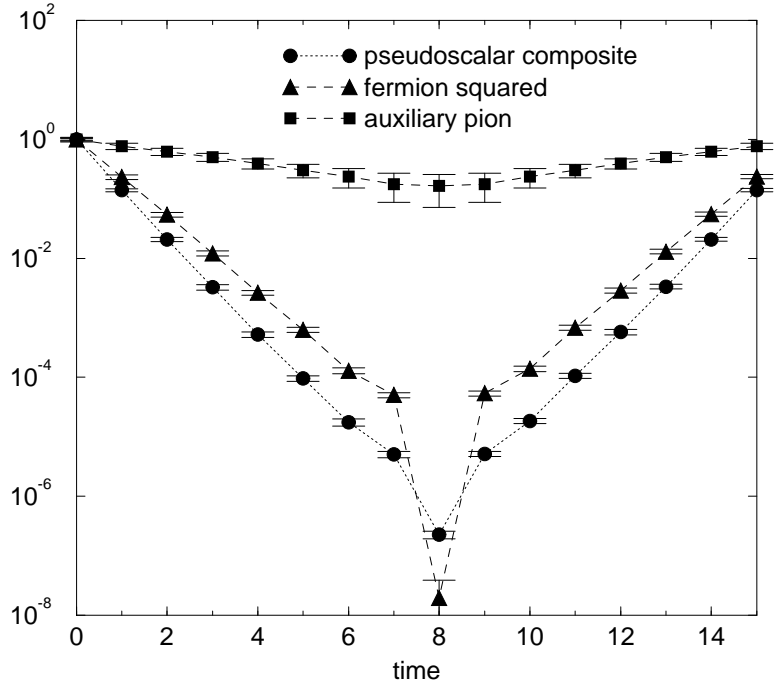,width=12cm}
\vspace{0.5cm}
\caption[xxx]{
Level ordering in 3d GN: from top to bottom,
auxiliary pion propagators, fermion propagators
squared, composite pseudoscalar propagator
at $\beta = 0.5$, $m_q = 0.01$, $N=3$ showing that
$m_{PS} > 2 m_f >> m_{\pi}$. }
\label{fig:gg}
\end{figure}

In Fig. \ref{fig:gg} we have plotted the propagator for the 
pion (measured from the auxiliary field), the square of the fermion
propagator and the pseudoscalar propagator. The results show that
 the pseudoscalar state is very close to
twice the physical fermion mass and $m_{PS} > 2m_f >> m_\pi$. 
What is remarkable is the fact that the 
auxiliary field pion is considerably lighter than $m_{PS}$. This can be 
understood in terms of the mechanism of chiral symmetry restoration in 
4-fermion interaction models whereby the disconnected contributions are 
responsible for making the pion light. We will provide numerical
evidence for this below. If we measure the pseudoscalar 
propagator in
the standard way from $GG^{\dagger}$ we include only the connected
contributions to the particle mass.

\begin{figure}
\vspace{-0.3cm}
\centerline{\epsfig{file=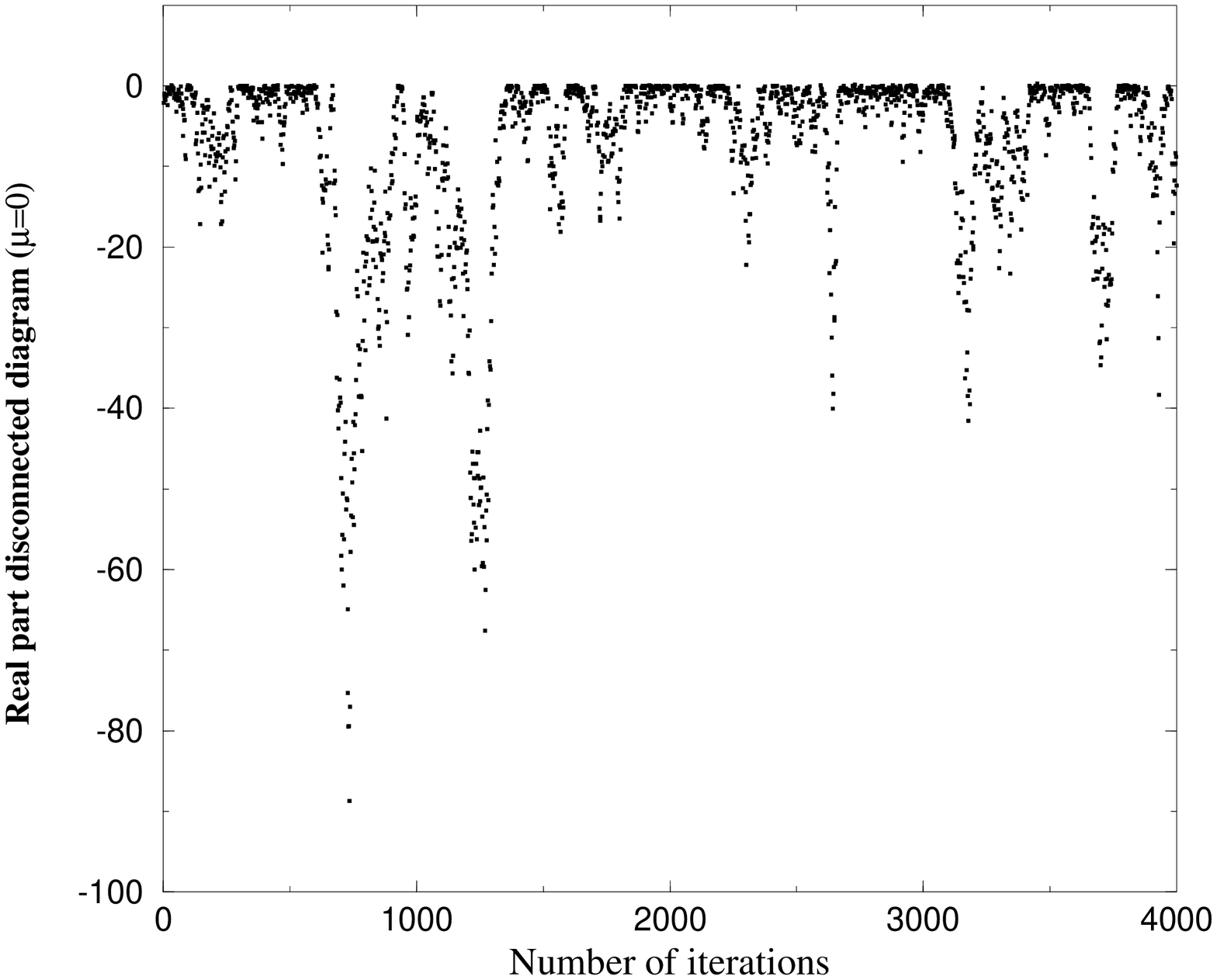,width=10cm,angle=-90,clip=}}
\vspace{0.2cm}
\centerline{\epsfig{file=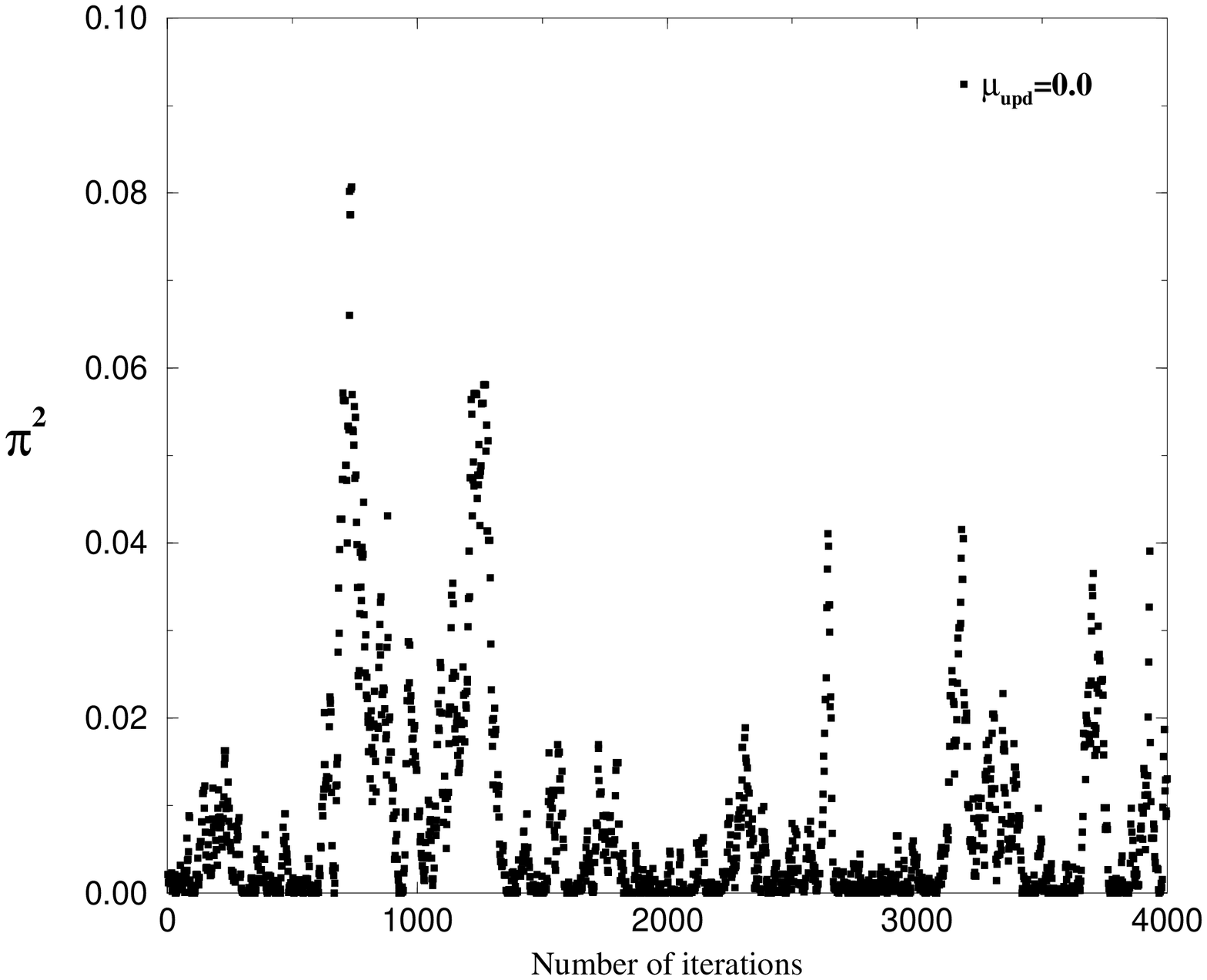,width=10cm,angle=-90,clip=}}
\vspace{0.3cm}
\caption{ Measurements at $\mu=0.0$  showing the disconnected 
contributions to the pion susceptibility (upper) on a $16^3$ lattice with $1/g^{2} =0.5$, $N=3$ and 
$m=0.01$ which we expect to be correlated with the auxiliary $\pi$ field via an 
equation of motion.}
\label{fig:mu0discon_pisq}
\end{figure}

\begin{figure}
\centerline{\epsfig{file=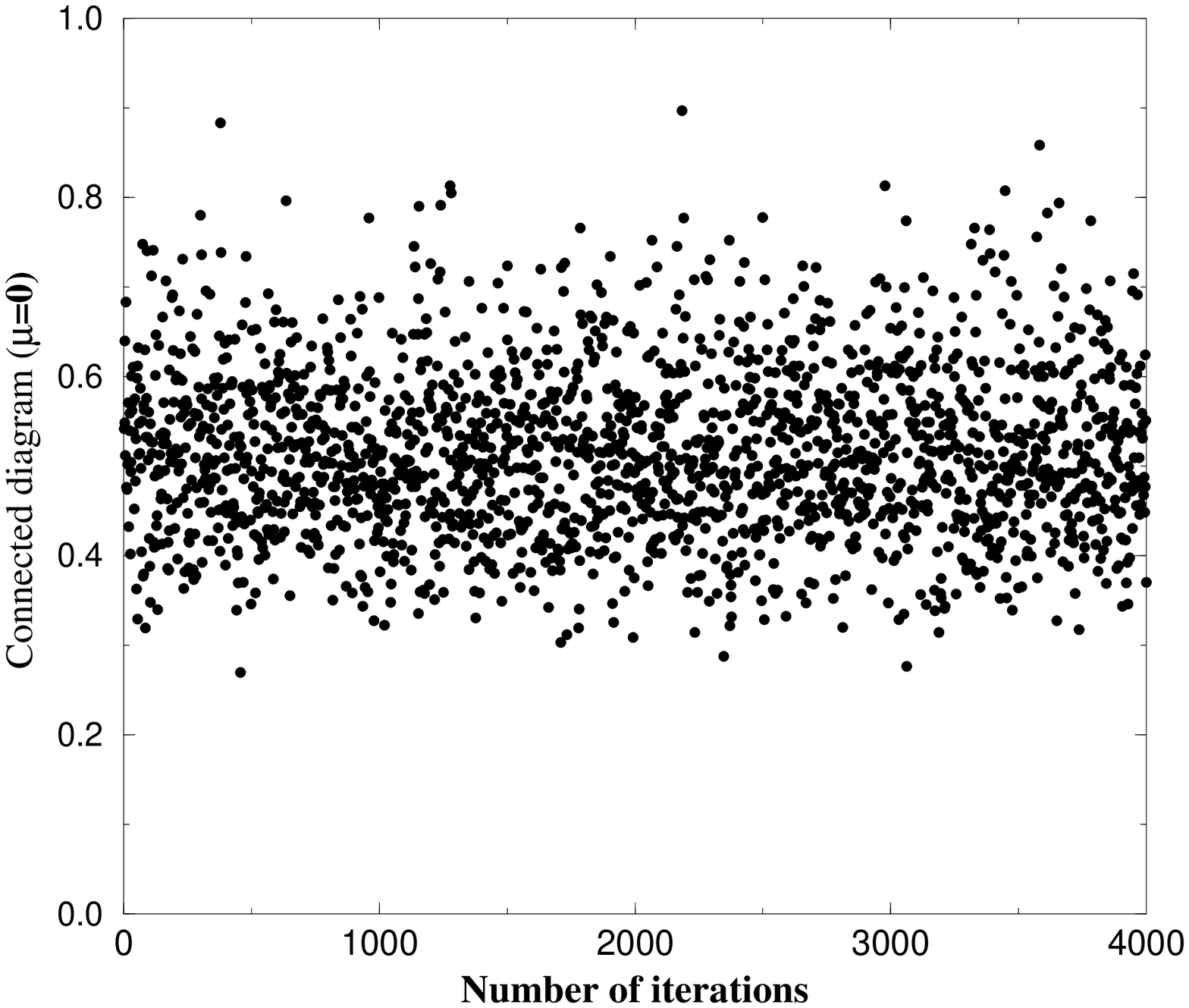,width=10cm,angle=-90,clip=}}
\vspace{0.2cm}
\centerline{\epsfig{file=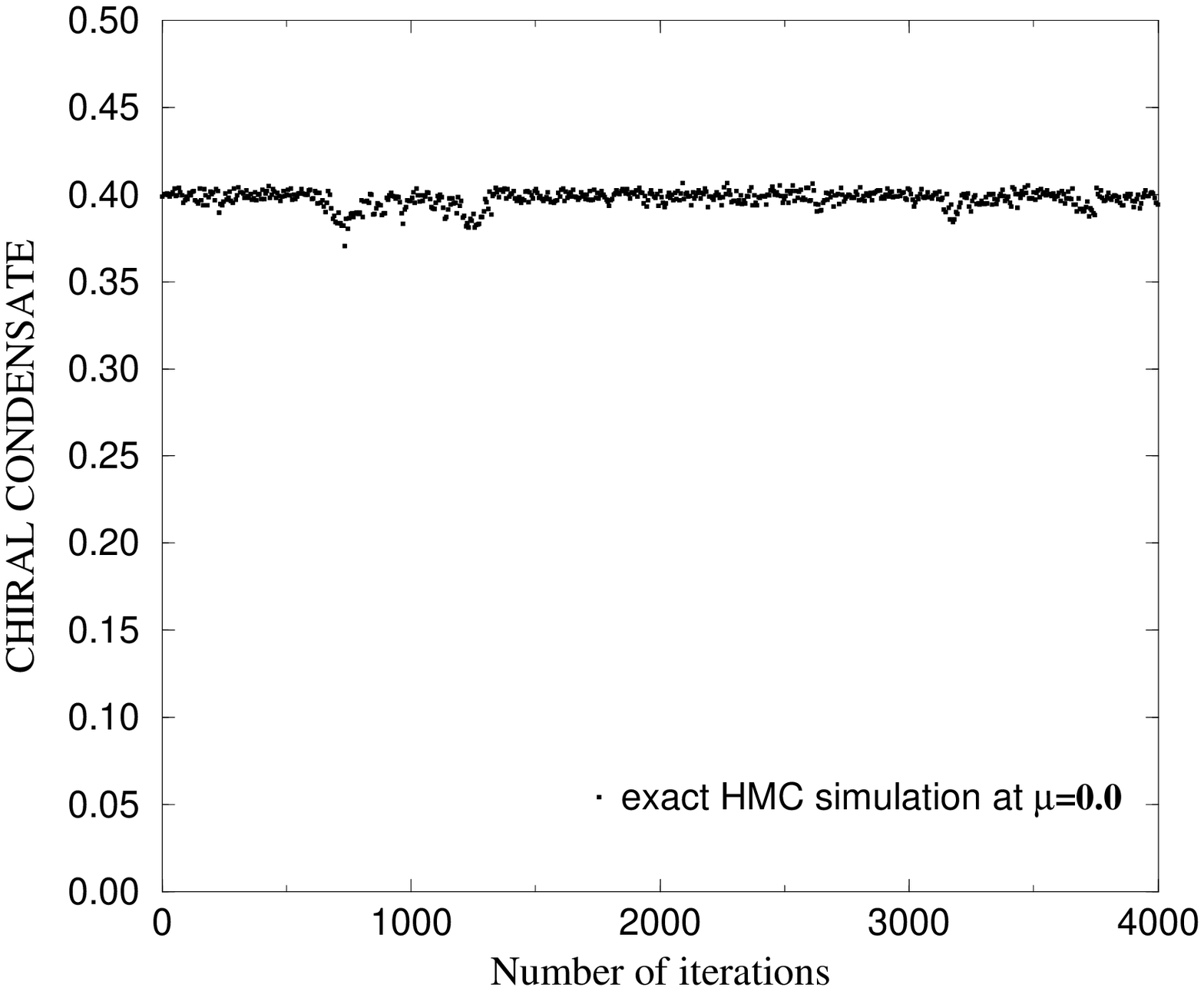,height=10cm}}
\vspace{0.5cm}
\caption{Connected contribution to the pion susceptibility (upper)
and the chiral condensate (lower) at $\mu=0.0$ on $16^3$ lattice with  
$1/g^{2} =0.5$, $N=3$ and $m=0.01$.}
\label{fig:mu0psibpsi}
\end{figure}

 Fig. \ref{fig:mu0discon_pisq} shows the disconnected
contribution to the Ward identity for a standard hybrid Monte Carlo 
simulation with $N=3$ at zero chemical potential. For this simulation we found that the signal for
chiral condensate from the noisy estimator was relatively stable as shown in
Fig. \ref{fig:mu0psibpsi} giving
${{\langle\psi\bar{\psi}\rangle}/{m_q}} \simeq 40$, therefore we require that the
sum of the connected and disconnected diagrams be of similar order
so that Eqn. \ref{eqn:ward} is satisfied. The connected contribution was 
stable at a value of around 0.5 therefore the dominant contribution
must come from the disconnected diagram. We found that the disconnected
contribution  was very noisy with large
downward peaks. The data suggests that considering
the connected contribution alone will never be sufficient to satisfy 
Eqn. \ref{eqn:ward}.  

 We repeated our measurements for
a non-zero chemical potential. We chose $\mu=0.5$ thus ensuring that we 
were still in the phase of broken chiral symmetry.
The results were consistent with those at zero chemical potential
as one would expect.

As a consistency check observe that we expect the
following relationships to hold from the equations of motion:
\begin{eqnarray}
 \bar{\psi}\psi = \frac{1}{g^2} \sigma \;\;\;;\;\;\;
\sum_{y}\bar{\psi}\varepsilon\psi(x)\bar{\psi}\varepsilon\psi(y)
= {\frac{1}{g^4}}{\frac{1}{V}}\left({\sum_{x}\pi(x)}\right)^2 
\end{eqnarray}

The first relationship has been checked and is satisfied by the simulation
data. The
second relation can also be verified. As seen in 
Fig \ref{fig:mu0discon_pisq}
the characteristic peaks in the disconnected contribution are
clearly correlated with the square of the pion field. 
We note that in the exact HMC study \cite{HKK3} the pion
mass was obtained from the correlator of the auxiliary field
$\pi$. 

The disconnected contribution to the pion susceptibility
gives a very noisy signal which suggests 
that a very long run would be required to equilibrate sufficiently 
to satisfy the lattice Ward identity. 

\section{Relevance of GN Study to QCD Simulations}

The Glasgow Method for QCD is very similar to the method described in 
section III of this paper however because the QCD action is complex for
($\mu\neq 0$)we are restricted to generating the ensemble at $\mu_{upd}=0$.
Furthermore the relationship $\det M{\hat{M}}=\det MM^{\dagger}$ which
we used in the lattice Gross-Neveu model does not hold for QCD at $\mu\neq0$. 
We have seen in section IV that the GN model observables are sensitive
to the choice of $\mu_{upd}$ and the most accurate results were obtained
for $\mu_{upd}\simeq \mu_c$. Clearly it is impossible to simulate QCD
at $\mu_{upd}\simeq m_{p}/3$ therefore we should be aware that very high 
statistics are likely to be required to obtain accurate observables from
the Glasgow Method for QCD. It is still possible that the Lee-Yang zeros 
distribution will give a hint for $\mu_c$ even with moderate statistics and 
$\mu_{upd}=0$. Since the Gross-Neveu model is a purely fermionic theory
it is impossible to assess from our study what influence the gauge fields have in
QCD dynamics in the context of Glasgow method reweighting \cite{SC}. 
 
In section V of this paper we have proved that the disconnected contribution
to the Goldstone pion is dominant in the Gross-Neveu model. In QCD
on the other hand we know that the pion has a dominant connected
contribution which implies that
$$
G_{ps}(t)\propto e^{-m_{\pi}t} 
$$
(c.f. eqn. \ref{GN_pion}) and this explains the origins of the early
onset of the chiral transition in QCD associated with a light baryonic 
pion. The existence of the baryonic pion in the Gross-Neveu model does
not lead to an early onset of the chiral transition because it is the 
disconnected contributions which make the pion light i.e. the Goldstone mechanism is realised in a fundamentally different way.

\section{Concluding Remarks}

In this paper we have found an example of a model where the inclusion of
the chemical potential into the dynamics is essential to obtain
exact results. 

We have proved that the physical observables of the Glasgow Method are
sensitive to the choice of $\mu_{upd}$. The optimal choice for revealing
critical behaviour is $\mu_{upd}\simeq\mu_c$, a choice which is impractical 
for QCD simulations. 
 In QCD the pseudoscalar channel pole is formed from {\it connected}
 diagrams corresponding to $G_{+\mu}(t)G^{\dagger}_{-\mu}(t)$. A 
  fermion pair with nonzero baryon charge (baryonic pion \cite{KLS}) forms from a 
 quark and a conjugate quark \cite{MISHA}
 (i.e. a fermion associated with $M^{\dagger}$ but with the sign of $\mu$ 
 reversed) and as a consequence the mass scale of the lightest baryon 
($m_{lb}$) is $m_{\pi}/2$ not $m_{p}/3$. 

This induces the 
 unphysical early onset of chiral symmetry restoration in quenched QCD.
In the full QCD it is conceivable that the phase of the determinant 
will eliminate the equality between quarks and conjugate quarks  to
allow us to recover the physical result $m_{lb}=m_{p}/3$. In principle the factor $R_{rw}$ in the Glasgow algorithm should influence the observables
 to take account of the phase of the determinant. The Glasgow method
QCD results \cite{LAT96}
still suggest $m_{lb}\simeq m_{\pi}/2$. This is likely to be explained by
ineffectiveness of the reweighting. The reweighting in the Gross-Neveu
model has been shown to be ineffective and this is not surprising because
in this case the statistical ensembles characterising the two phases are 
non-overlapping. The Glasgow method may however be more effective in 
models more sophisticated than Gross-Neveu and in other dynamical regimes such
as at high temperature.

In 3d GN $U(1)$ the Goldstone mechanism is realised by a
pseudoscalar channel pole formed from {\it disconnected} diagrams and 
the state  $G_{+\mu}(t)G^{\dagger}_{-\mu}(t)$ yields a bound state
of mass $2m_f(\mu=0)$ which is considerably heavier than the pion. 
As a consequence even when we consider 
individual configurations in this model we find 
$\mu_o\simeq\mu_c\gg m_{\pi}/2$.

The different realisations of chiral symmetry breaking in the GN model and QCD
have a simple physical origin. For QCD, a vector-like theory, like charges
repel, so that light states only form between $q\bar q$ pairs 
(the pathologies in
simulations of QCD with $\mu\not=0$ occur because of the influence of
conjugate quarks in the measure; in this case light states may be formed
from $qq^c$ pairs). In theories with Yukawa-like interactions such as the GN
model, however, in the Born approximation all interactions are attractive, and
thus one might expect light states made of $q\bar q$, $qq$ and $\bar q\bar q$.
Only the $q\bar q$ system has contributions from the disconnected diagram,
however. As we have seen,
binding in the connected channel is insufficient to make light
states. Therefore we see a natural relation between the dominance of the
disconnected diagram and the absence of light states made of two quarks in the
spectrum. 

It is also worth reconsidering why simulations of the GN model seem not to be
afflicted with the problems observed in QCD. As we saw in Section
{\ref{sec:lat}}, the price of simulating with a real measure
$\mbox{det}(M^\dagger M)$ is that the model contains both ``white'' and ``black''
fermions with opposite axial charges. We might therefore worry about
attractive interactions between white and black fermions
and the possibility of
light bound state in the $G_{+\mu}^{white}G_{+\mu}^{black}$ channel
The reason that no spurious
onset at $\mu=m_\pi/2$ occurs, and that the
GN simulations yield results in good agreement with the mean field predictions,
is intimately related to the fact that no light state occurs in this channel
indeed the white-black interaction due to pion exchange in
this model has the opposite sign and is thus
{\sl repulsive\/}:
this is in contrast to the situation in gauge theories with real
measure, where
the interaction between quarks and conjugate quarks is attractive.

There are interesting parallels with the standard discussion of the Vafa-Witten
theorem \cite{VW}, forbidding the spontaneous breaking of global vectorlike
symmetries such as isospin or baryon number in field theories with a
positive definite measure in the path integral (this includes gauge theories
with zero chemical potential but {\sl not\/} Yukawa or GN models, for which
the measure is in general complex). Vafa and Witten do in fact discuss
a `white/black' model with a real measure similar to ours,
but with Yukawa couplings to a
pseudoscalar field only. Their analysis does apply in this case, and a
flavor-violating $\langle\bar\psi^{white}\psi^{black}\rangle$ condensate
is forbidden, the reason being simply that the interaction between white and black
particles is repulsive.

Once Yukawa couplings to scalar degrees of freedom are introduced, however,
as in the GN model, the inequalities necessary to prove the Vafa-Witten theorem
no longer hold. There appears to be no fundamental obstruction, therefore, to
the generation of a baryon-number violating diquark condensate
$\langle\psi^{white}\psi^{white}\rangle$ in this model,
despite its measure being real;
it has been recently suggested that such a condensate forms in QCD at high
density \cite{DIQ}. A lattice study of this phenomenon is in progress
\cite{us}.

\section{Acknowledgements}
We thank J.F. Laga{\"e} for his comments related to measurement of 
the pion mass. MPL, SEM and SJH received support from EU TMR contract no. ERBFMRXCT97-0122. JBK thanks
the National Science Foundation ( NSF-PHY96-05199 ) for partial support.

\newpage
\newpage

\newpage

\newpage

\newpage

\newpage
\begin{figure}
\centerline{{\epsfig{file=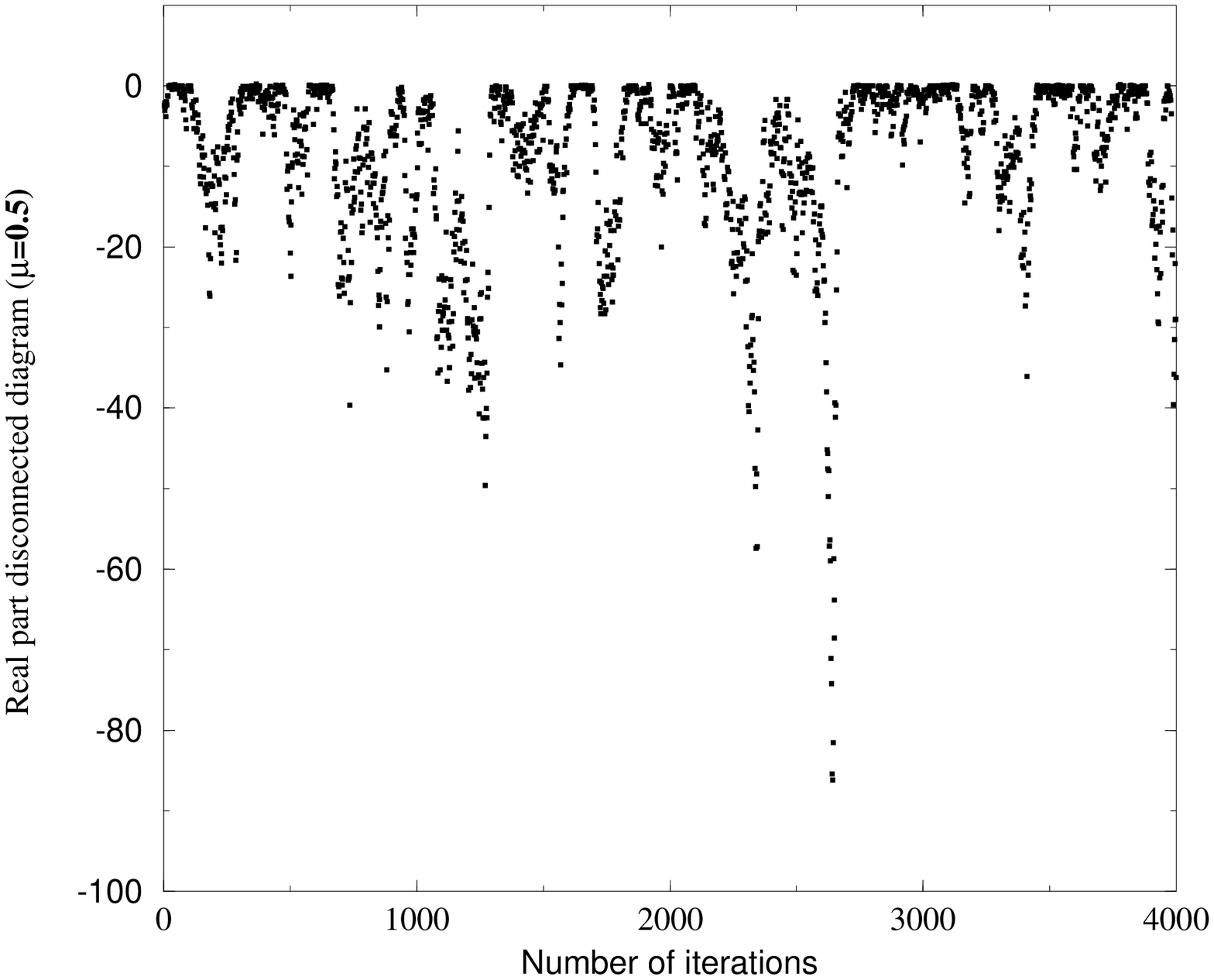,height=10cm,angle=-90}}}
\vspace{0.4cm}
\centerline{{\epsfig{file=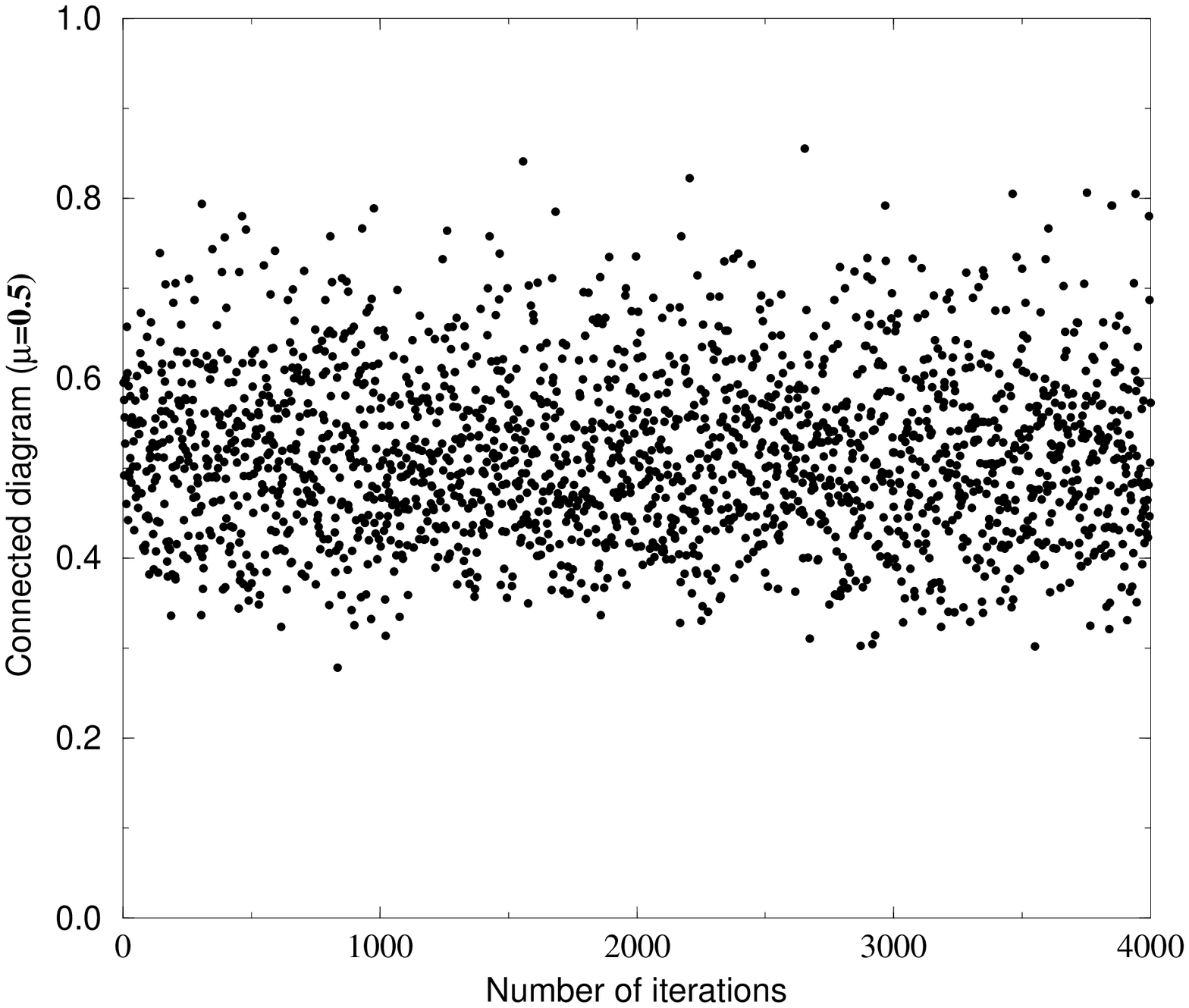,height=10cm,angle=-90}}}
\vspace{0.5cm}
\caption{Measurements at $\mu=0.5$ showing the relative magnitudes
of the disconnected (top) and the connected (bottom) contribution to the pion
on a $16^3$ lattice with $1/g^{2} =0.5$, $N=3$ and $m=0.01$.}
\label{fig:mu0.5_con_discon}
\end{figure}

\begin{references}
\bibitem{OLD} I. Barbour {\it et.al.},Nucl. Phys. {\bf B275},
296 (1986).
\bibitem{DAVKLEP} C.T.H. Davies and E.G. Klepfish Phys. Lett. {\bf 256B},
68 (1991).
\bibitem{MISHA} M. Stephanov, Phys. Rev. Lett. {\bf 76}, 4472 (1996).
\bibitem {Go88} A.~Gocksch, Phys. Rev. {\bf D37} (1988) 1014.
\bibitem{KLS} 
M.-P. Lombardo, J.B. Kogut, D.K. Sinclair, Phys. Rev. {\bf D54},2303 (1996).
\bibitem{HKK3} S. Hands, S. Kim, J.B. Kogut, Nucl. Phys. {\bf B442},
364 (1995).
\bibitem{MISHA2} M.A. Halasz, A.D. Jackson, R.E. Shrock, M.A. Stephanov,
J.J.M. Verbaarschot, hep-ph/9804290
\bibitem{BARBELL} I.M. Barbour, A.J. Bell, Nucl. Phys. {\bf B372},
385 (1992).
\bibitem{LAT96}I.M. Barbour, J.B. Kogut, S.E. Morrison, Nucl. Phys. B
(Proc. Suppl.) {\bf B53}, 456 (1997).
\bibitem{BILIC} N. Bili\'c, K. Demeterfi, B. Petersson,
Nucl. Phys. {\bf B377} 615 (1992).
\bibitem{SC} I. Barbour, S.E. Morrison, E. Klepfish, J.B. Kogut, M.P. Lombardo,
 Phys. Rev. {\bf D56},7063 (1997).
\bibitem {AT92} A.~Hasenfratz and D.~Toussaint, Nucl. Phys. {\bf B371}
(1992) 539
\bibitem{DKPR} S. Duane, A.D. Kennedy, B.J. Pendleton and D. Roweth,
Phys. Lett. {\bf B195}, 216 (1987).
\bibitem{BB} C.J. Burden and A.N. Burkitt, Europhys. Lett. {\bf3}, 545 (1987)
\bibitem{HK} S.J. Hands and J.B. Kogut, hep-lat/9705015. (1997).
\bibitem{GIBBS2} P. Gibbs, Phys. Lett. {\bf B172}, 53 (1986); P. Gibbs, Phys. Lett. {\bf B182}, 369 (1986).
\bibitem{BIELWSP} Susan E. Morrison, Nucl. Phys. {\bf A642}, 269 (1998).
\bibitem{MATSTONE84} H. Matsuoka, M. Stone, Phys. Lett. {\bf 136B},
204 (1984).
\bibitem{LEEYANG} T.D. Lee, C.N. Yang, Phys. Rev. 
{\bf 87}, 404 (1952); Phys. Rev. {\bf 87}, 410 (1952).
\bibitem{HKK_AP} 
S. Hands, A. Kocic, J.B. Kogut, Ann. Phys. {\bf 224}, (1993) 29.
\bibitem{RWP} B. Rosenstein, B.J. Warr and S.H. Park, Phys. Rev. {\bf D39},
3088 (1989).
\bibitem{HKK4} S.J. Hands, A. Koci\'c and J.B. Kogut, Nucl. Phys. {\bf B390},
355 (1993).
\bibitem{VW} C. Vafa and E. Witten, Nucl. Phys. {\bf B234}, 173 (1984).
\bibitem{DIQ} 
M. Alford, K. Rajagopal, F. Wilczek, Phys. Lett. {\bf B422}, 247 (1998);
R. Rapp, T. Schaefer, E.V. Shuryak, M. Velkovsky, Phys. Rev. Lett. {\bf 81}, 
53, (1998).
\bibitem{us} S.J. Hands and S.E. Morrison, hep-lat/9807033.
%
\end{references}
\end{document}